\documentclass[twocolumn]{revtex4}
\usepackage{epsfig,latexsym}
\usepackage{amsmath,amscd,amssymb}
\usepackage{bm}
\usepackage{graphicx}
\usepackage{color}

\def\kB{k_{\rm B}}
\def\NR{N_{\rm cyl.}}
\def\NB{N_{\rm con.}}
\def\xR{x_{\rm cyl.}}

\begin{document}
\title{Clusters of proteins in bio-membranes: insights into the \\ roles of interaction potential shapes and of protein diversity}
\author{Nicolas Meilhac$^{\dagger\ddagger}$, Nicolas Destainville$^{\dagger\ddagger}$} 

\affiliation{$^{\dagger}$Universit\'e de Toulouse; UPS; Laboratoire de
Physique Th\'eorique (IRSAMC); F-31062 Toulouse, France\\
$^{\ddagger}$CNRS; LPT (IRSAMC); F-31062 Toulouse, France}

\date{\today}

\begin{abstract}
It has recently been proposed that proteins embedded in lipidic bio-membranes can spontaneously self-organize into stable small clusters, or membrane nano-domains, due to the competition between short-range attractive and longer-range repulsive forces between proteins, specific to these systems. In this paper, we carry on our investigation, by Monte Carlo simulations, of different aspects of cluster phases of proteins in bio-membranes. First, we compare different long-range potentials (including notably three-body terms) to demonstrate that the existence of cluster phases should be quite generic. Furthermore, a real membrane contains hundreds of different protein species that are far from being randomly distributed in these nano-domains. We take this protein diversity into account by modulating protein-protein interaction potentials both at short and longer range. We confirm theoretical predictions in terms of biological cluster specialization by deciphering how clusters recruit only a few protein species. In this respect, we highlight that cluster phases can turn out to be an advantage at the biological level, for example by enhancing the cell response to external stimuli. 
\end{abstract}

\maketitle

\section{Introduction}

Live cells contain a large amount of membranes playing a great variety of biological roles. Membranes are an essential component of Life. Their first function is to separate the diverse cell organelles (e.g. the Golgi apparatus, mitochondria, the endoplasmic reticulum, the nucleus in animal cells) from each other. The plasma membrane, on which this paper primarily focusses, separates the interior of the cell from the environment. However, membranes are not passive barriers but they control in an active way the movement of molecules or information in and out the organelles or the cell, thus maintaining substantial differences of composition between the cell and its environment. Plasma membranes also play an important role in cell adhesion and motility, thus ensuring a large amount of biological functions. The principal constituents of plasma membranes are lipids and proteins, accounting each for typically 50~\% of the membrane mass, and constituting a two-dimensional complex fluid where interactions are weak, typically on the order of the thermal energy $\kB T$~\cite{Alberts,Phillips}. Adopting a coarse-grained point of view, lipids can be considered as an underlying fluid (or solvent) in which included proteins evolve. Effective protein-protein interactions mediated by this fluid emerge, because an inclusion creates a perturbation of the membrane that in turn influences neighbor inclusions. 
%[citer Nature Phillips 2009]. NON: FIG. FAUSSE

Membrane organization is a topical issue in cell biology. Understanding how hundreds of different protein and lipid species organize in a same membrane to perform as many biological functions~\cite{Alberts,Dowhan} remains a challenge to which physicists have recently paid attention, by applying concepts of soft matter to the biophysics of bio-membranes~\cite{Phillips,Lenne09}. Far from the pioneer mosaic model of Singer and Nicolson~\cite{Singer}, it is now understood that lipids and proteins are non-homogeneously distributed in membranes. 
They are subtly organized in small domains, or compartments, or clusters~\cite{Lang10}, of highly variable composition, the length-scale of which ranges from a few nanometers to microns. Heterogeneity is even thought to be a key ingredient for biological functions by concentrating a few protein species in a same nano-domain. Indeed, it facilitates the encounter of different proteins species that must interact to perform together a given biological task~\cite{Simons97,Davare01,Laporte01,Rozenfeld10}. When in a same membrane domain, they encounter much more easily than if they were randomly diffusing on the membrane surface~\cite{Simons00}. Concentrating identical receptors in a same domain also facilitates the response to small inputs by reducing noise~\cite{Gurry09}. Grouping together structure proteins is also essential in the context of cell adhesion~\cite{Espenel08,Cavey08}. A wide range of experimental techniques demonstrate that co-localization of a few different protein species in the same membrane nano-domains is a common feature (see, e.g.,~\cite{Davare01,Prior01B,Koopman04,Park05,Sieber06,Sieber07,Hess07,Abankwa07,Prior07,deBakker08,Espenel08,Goswani08,Gurry09,Grage11}) and as stated by T. Lang and S.O. Rozzoli, clusters are ``more than a pretty picture''~\cite{Lang10}.

However, the precise way proteins organize and the physicochemical mechanisms responsible for this organization are currently a matter of controversy~\cite{Heerklotz02,Daumas03,Chen04,Sieber06,Poveda08,Gurry09,Lingwood10}. In particular, a consensus has not been reached yet on the reason why domains remain sub-micrometric. The ideas developed in this paper propose elements of answers to the
two following questions: 1) What causes compartmentalization of membrane proteins in sub-micrometric domains  in live cells?  2) How can compartmentalization mechanisms account for segregation of membrane proteins in specialized, heterogeneous sub-micrometric structures where they more easily encounter to perform biological functions? 

Motivated by this biophysical context, statistical mechanics arguments have recently been proposed to account for the existence of protein nano- or micro-domains in a membrane at equilibrium~\cite{Daumas03,Sieber07,Gurry09,Evans03,PRE08A,PRE08B}. They rely on the following mechanism: attraction at short range ($\sim~1$~nm) between proteins favors condensation of membrane proteins in a dense phase, but some weaker repulsion at longer range ($\sim~10$~nm) prevents a complete phase separation because once clusters have nucleated, 
sufficiently large clusters repel each other due to additivity of the repulsion between their proteins. Alternatively, too large clusters are unstable because of the long-range repulsion. The resulting phase at equilibrium is called a ``cluster phase''~\cite{PRE08A}, by analogy with similar phases in soft condensed matter~\cite{Stradner,Seul95,Giulani08}. However, we will show that real protein-protein interaction potentials are very complex and potentially very diverse. The present work addresses the question of the robustness of the cluster phases with respect to the potential shape and demonstrates that cluster phases should be generic in bio-membranes in spite of potential diversity.

In addition, previous studies did not take into account protein diversity. Recent theoretical investigations have shown that modulating the short-range attraction in order to account for diversity could lead to composition heterogeneity of clusters because the ensuing energy gain is larger than the corresponding entropic cost~\cite{EPL10}. The present work also addresses this question on numerical grounds and confirms these theoretical investigations (see also~\cite{deMeyer10}, Supporting Material).

The paper is thus organized as follows. We recall the biophysical context and we present our model in the two first sections. Then Section 4 is mainly devoted to the question of the robustness of cluster phases with respect to the potential shape, more precisely to its long-range repulsion term. In Section 5, we focus on the role of protein diversity through its effects on the short-range attraction. The last section is devoted to discussions and conclusions.

\section{Biophysical context}
\label{sota}

Interactions between proteins embedded in bio-membranes are manifold and the literature on the topic is extensive. In addition to hard-core and electrostatic interactions, they feel several effective forces specific to these systems, because they are mediated by the membrane. As we shall see, the free energies involved in these interactions are on the order of the thermal energy $\kB T$ and thus play a role at equilibrium. 

If the membrane is seen as an elastic sheet of curvature elastic modulus $K$, it is locally deformed by the presence of a protein, whether it be embedded or peripheral, and the response to this constraint implies effective  long-range forces. First, thermal undulations of the elastic membrane in the dimension normal to the membrane plane are perturbed by the presence of proteins, in a way that depends on their separation $r$. The entropy,
depending on the number of effectively accessible undulation modes, thus depends on separation, and an attractive Casimir-like force ensues, which is pairwise additive and decays as $r^{-4}$ at large $r$~\cite{Goulian93,Fournier97}. In addition, when proteins are (up-down) asymmetric with respect to the membrane plane, which we assume to be the majority case for both peripheral and transmembrane proteins, they can be modeled as conical inclusions with a half-aperture angle (or contact angle) denoted by $\alpha$. This angle models not only the crystallographic shape of the protein~\cite{Chou01,Doyle98}, but also its local interactions, in particular of electrostatic nature, with the surrounding lipids. The two lipid monolayers of cell membranes are very different in composition and in charge~\cite{Alberts}, which also breaks the  
up-down symmetry, even for cylindrical inclusions. Such asymmetric inclusions feel an additional repulsion, also decaying as $Kr^{-4}$ at leading order. This repulsion, proportional to the squares of the effective contact angles $\alpha$, is due to the elastic deformation imposed to the elastic membrane~\cite{Goulian93,Fournier97,Kim98,Dommersnes99,Kim99}.  Note the important following point: even for differently oriented conical inclusions with contact angles $\alpha_1$ and $\alpha_2$ of {\em opposite sign}, this elastic contribution remains repulsive because it is proportionnal to $\alpha_1^2+\alpha_2^2$ at leading order. This repulsion compensates the Casimir-like attraction as soon as $\alpha$ is typically larger than 5$^\circ$. It is not pairwise additive anymore, since three-body terms, which can be attractive, exist at the leading order $r^{-4}$~\cite{Dommersnes99,Fournier03,Kim99}. But we will demonstrate that they are not strong enough to counter-balance repulsive two-body forces at long range (see also the Supporting Information). As for $n$-body interactions terms with $n>3$, they decay faster than $r^{-4}$~\cite{Dommersnes99} and will not be considered in the present work~\cite{Kim99}. 
Furthermore, proteins need not be isotropic inclusions, in this sense that their horizontal section can be better modeled by an ellipse (instead of a disk). In this case, the potential depends on the relative orientation of the ellipses~\cite{Dommersnes99,Chou01}, and decays as $r^{-2}$. But except for strong anisotropies, it is averaged over orientations because of the rapid protein rotational diffusion~\cite{Chou01}, which significantly lowers the strength of the interaction and makes it decaying as $r^{-4}$ as well. In case of strong anisotropy, the ensuing elastic interaction can even become attractive~\cite{Chou01}. In the present work, we only consider weakly or moderately anisotropic inclusions, such that the net elastic interaction is repulsive at long range. We discuss strong anisotropy in the Supporting Information. Beyond anisotropy, protein shape can make interaction potentials even more complex~\cite{Fournier99}.

Even though these interactions mediated by the membrane should decay algebraically at long range, they are likely to be partially screened for at least two reasons: 
%High protein density leads to an effective membrane softening at long range~\cite{Netz95}; NON : calculs ˆ petit \delta \kappa alors que nous sommes ˆ la limite des grands \delta \kappa
1) Long wavelength membrane excitations are damped when the membrane is coupled to the much more rigid cytoskeleton~\cite{Lorman06}; 2) When weak, but non-vanishing membrane tension is taken into account, the repulsion is screened beyond distances of a few tens of nanometers and thus decays exponentially at long range~\cite{Weikl98} (see also~\cite{Evans03,Manghi06} for related calculations). However, estimating the values of membrane tensions in live cells, and therefore of the screening length, is a difficult experimental task. It is thus relevant to consider both exponentially and algebraically decaying potentials.

At shorter range, proteins first experience a hard-core repulsion at contact. In this respect, protein radii $a$ are variable, ranging from a fraction of nanometer for single membrane-spanning $\alpha$-helices, to a couple of nanometers for large proteins (such as receptors, pumps or channels)~\cite{Alberts}. As a first step, we consider mono-disperse distributions of protein radii $a$ in this work, with the typical radius $a=2$~nm. The general case is briefly discussed in the Supporting Information. At contact or close to contact, interactions are of hydrophobic, or (screened) electrostatic, or hydrogen-bond origins, arising from apolar, or polar, or charged amino-acids near the protein surface. Because of the rotational averaging discussed above, we do not take into account the anisotropy of these interactions either. 
Van der Waals forces (in $r^{-6}$) should also play a role at short distances. 

Two additional forces mediated by the lipidic membrane play a role when separations between protein surfaces are typically less than one nanometer. The hydrophobic core of transmembrane proteins is likely not to match the equilibrium membrane thickness, what is usually called hydrophobic mismatch. The mismatch is said to be positive when the core is too thick, and negative in the converse case. Two proteins with mismatches of identical (resp. different) sign feel an attractive (resp. repulsive) short-range force due to the induced elastic variations of the bilayer thickness~\cite{2,Nielsen98,Bohinc03,deMeyer08,Grage11}. Mismatch has been experimentally demonstrated to promote the aggregation of transmembrane proteins~\cite{Botelho06}. There also exists many-body effects associated with hydrophobic mismatch~\cite{Brannigan07}. They are not included in the present study because no simple effective expression is known for these forces. 

In addition, attractive depletion forces, due to the two-dimensional osmotic pressure of lipids on proteins, tend to bring them closer when they are about a nanometer apart~\cite{19}. Hydrophobic and depletion forces have been studied numerically by Molecular Dynamics~\cite{Thogersen08,Weiss08,deMeyer08,West09,deMeyer10}. Their range is nanometric and the involved energies are equal to a few $\kB T$. Attractive or repulsive hydrophobic mismatch modulates this strength by a couple of $ \kB T$ and can consequently change the degree of aggregation (or oligomerization) of proteins (e.g. rhodopsins in~\cite{Periole07}).

Finally, cell membranes are constituted of a large variety of different lipid species~\cite{Hinderliter04} and proteins are known to recruit
in their neighborhood lipids for which they have a better affinity~\cite{Dowhan}. For two identical proteins, the closer they are, the more energetically favorable the
configuration because there is an interface energy associated with this ``wetting'' phenomenon. Typical energies are also of order $\kB T$~\cite{2,Hinderliter04,Gil98}. These references propose a protein-driven mechanism for domain formation invoking such forces, but the limited
domain size due to additional repulsive forces has not been
discussed in this context. 

To conclude, protein interaction potentials have diverse contributions, and writing a generic potential shape is a tedious task appealing to intensive Molecular Dynamics simulations, out of the scope of the present work. For this reason, it is useful to consider different realistic, typical shapes, and to analyze the similarities and the differences between the ensuing cluster phases, as already discussed in reference~\cite{PRE08A}. We focus on two types of interaction potentials:~potentials decaying algebraically as $r^{-4}$ at long range, corresponding to unscreened or partially screened repulsion; and potentials decaying exponentially at long range, corresponding to complete screening. We thus span a large variety of experimental contexts. We demonstrate that conclusions are qualitatively similar in both cases, thus proving that cluster phases of membrane proteins should be generic. 

\section{Model}

Here and in the following, all energies are implicitly in units of $\kB T$, because it is the relevant scale of energies in this context. 

The pairwise potentials $U_2(r)$ considered in this work comprise a hard-core repulsion at very short distances, i.e. $U_2(r)=\infty$ if $r<2a$, an attractive part at short range and a weak repulsive one at larger distances. The attractive part is chosen to be exponential in all cases, because the short-range forces considered above decay rapidly by nature. 
%Long-range potentials mediated by the membrane should decay algebraically, as $r^{-4}$. However, it is likely that they are at least partially screened because of high protein density~\cite{Netz95}, or damping of long wavelength membrane excitations when it is coupled to the much more rigid cytoskeleton~\cite{Lorman06}. It is thus relevant to consider potentials decaying exponentially at long range. They correspond to the extreme case of complete screening.

\begin{figure}[ht]
\begin{center}
\includegraphics[width=0.45\textwidth]{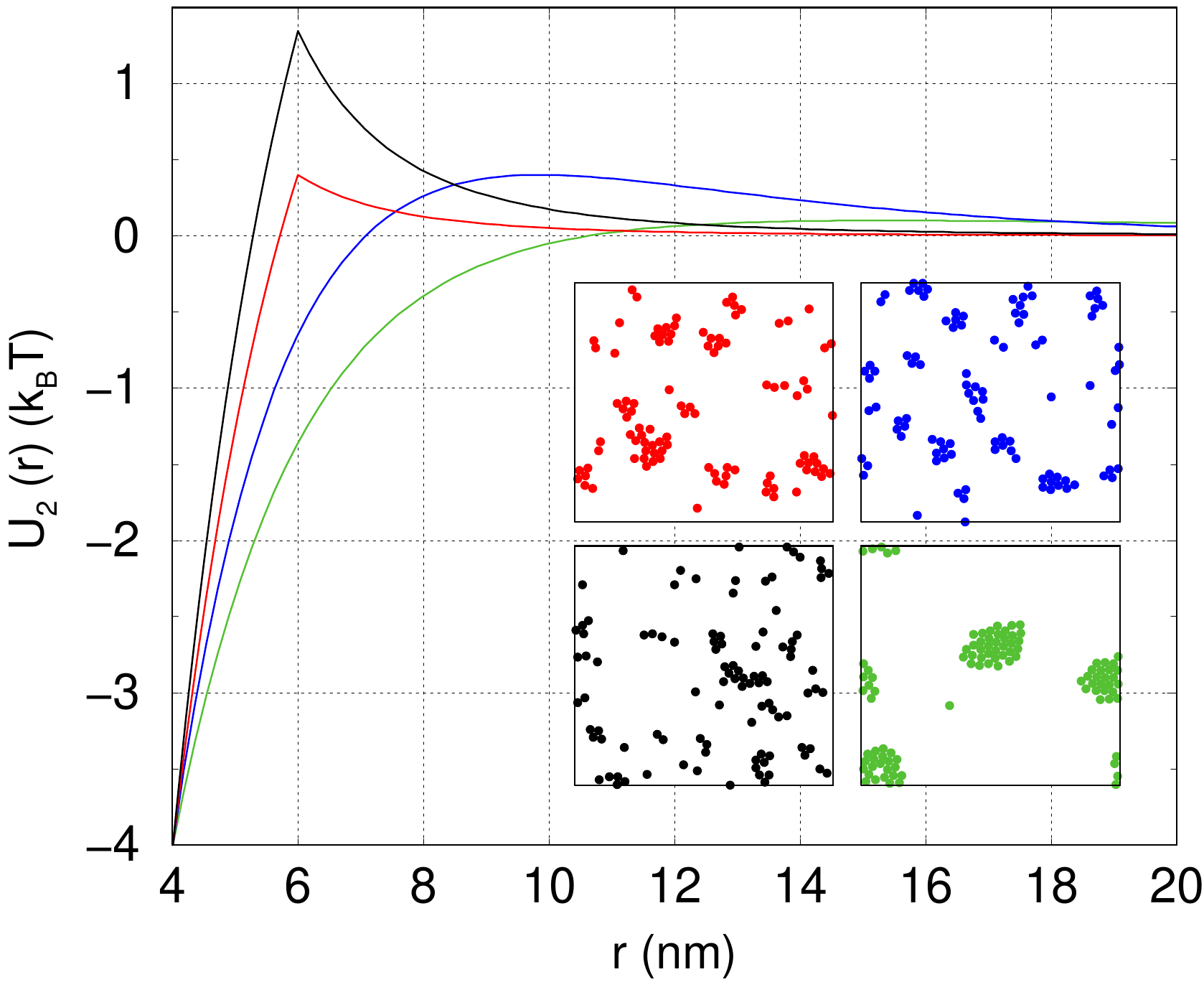}   
\caption{Examples of pair potentials $U_2$ used throughout this paper. 
From left to right at small $r$: two algebraic potentials [equation~(\ref{repulse})] with $K=150$ (black) and $50k_{\rm B}T$ (red); and two bi-exponential potentials [equation~(\ref{biexp})] with $E_{\rm max}=0.4$ (blue) and $0.1 k_{\rm B}T$ (green) and $1/\gamma_r=4$ (blue)  and 16~nm (green). $E_{\rm min}=-4 k_{\rm B}T$, and $1/\gamma_a=2$~nm in all cases. In all cases, cluster co-exist with monomers at equilibrium, as illustrated by the four random snapshots (at density $\phi=0.1$, same colors as the corresponding curves).
\label{exs0}}
\end{center}
\end{figure}

As motivated above, we will study on the one hand bi-exponential potentials of the form~\cite{PRE08A,Sear99,Imperio07}:
\begin{equation}
U_2^{\rm exp}(r)=-\varepsilon_a \exp(-\gamma_a r)+\varepsilon_r \exp(-\gamma_r r),
\label{biexp}
\end{equation}
where all parameters are positive. They contain an attractive and a repulsive term and $\gamma_r<\gamma_a$ (see examples in figure~\ref{exs0}). They correspond to complete screening. Parameter values will be specialized in our numerical studies, in agreement with the biophysical context of the previous Section. As in reference~\cite{PRE08A}, we set $1/\gamma_a=2$~nm, $1/\gamma_r=4$ or 16~nm, and $\varepsilon_a$ and $\varepsilon_r$ are adapted to fit the desired energy at contact $E_{\rm min}\equiv U_2^{\rm exp}(2a)$ and the desired energy barrier, $E_{\rm max}$. For example, $1/\gamma_r=4$~nm, $E_{\rm min}=-4 \kB T$ and $E_{\rm max}=0.4 \kB T$ lead to $\varepsilon_a = 55.1 \kB T$ and $\varepsilon_r = 9.4\kB T$. The maximum value $E_{\rm max}$ is then reached at $r=9.8$~nm.

On the other hand, we will consider unscreened potentials, which still display an exponential short-range attraction but the repulsive part of which is slowly, algebraically decaying:
\begin{equation}
U_2^{\rm pow}(r) = \left\{
\begin{array}{lcl}
 -\varepsilon_a \exp(-\gamma_a r) +E_s & \mbox{ if } & r \leq r_l  \\
 \displaystyle{\frac{C}{r^p} \phantom{+E_s}}   Ê&\mbox{ if } &r \geq r_l 
 \end{array}
\right.,
\label{repulse}
\end{equation}
where the energy shift $E_s$ enforces the continuity of $U$ at $r_l$. The existence of cluster phases for $p=2$ has been explored in reference~\cite{PRE08A}. They also exist for $p=6$ (data not shown). Here we focus on $p=4$ because the repulsion between isotropic inclusions in an elastic membrane displays such a long-range behavior, with
$C=S^2[\frac{4K}{\pi}(\alpha_1^2+\alpha_2^2)-\frac{6}{\pi^2}]$, where $S=\pi a^2$ is the inclusion area, $K$ is the membrane elastic modulus and $\alpha_{1,2}$ are contact angles~\cite{Goulian93,Fournier97}. The first term in the squared brackets corresponds to pure elastic deformations of the membrane and the second one to the Casimir-like forces.  The constant $C$ is positive for sufficiently large values of $K$ and $\alpha_{1,2}$. The values of the contact angles $\alpha$ are difficult to estimate experimentally. Here, we shall consider two typical values, $\alpha=0$ or $10^\circ$, so that, with the values of $K$ considered below, $C$ is always positive except if $\alpha_1=\alpha_2=0$. As above, we set $1/\gamma_a=2$~nm and we choose $r_l = 2a + 1/\gamma_a$, $1/\gamma_a$ being the typical extend of the attraction beyond the hard core distance, $2a$. Finally, $\varepsilon_a$ is set so that $E_{\rm min}$ has the desired value (see below). Examples are given in figure~\ref{exs0}.

As far as the elastic modulus $K$ is concerned, the experimental values found in the literature are very variable, because they depend strongly on the lipidic composition, in particular the cholesterol concentration. The typical values of $K$ range from 10 to $200 \kB T$~(\cite{Duwe90,Meleard97,Manneville99} and references therein). In this work, we consider the values $K=30$, 50 and $150 \kB T$, thus spanning the experimental interval. Accordingly, if $\alpha_1=\alpha_2=10^\circ$, then $E_{\rm max}=0.21$, 0.40 and $1.34 \kB T$, respectively.

In Section~\ref{intermediate}, we will also explore the role of long-range three-body forces mediated by the elastic membrane, also decaying as $r^{-4}$:
\begin{equation}
U_3^{\rm pow}=B \sum_{p<q<r}\left[ \frac{\cos(2\gamma_p)}{r_{pq}^2 r_{pr}^2}+\frac{\cos(2\gamma_q)}{r_{pq}^2 r_{qr}^2}+\frac{\cos(2\gamma_r)}{r_{pr}^2 r_{qr}^2} \right],
\label{3b}
\end{equation}
where $\gamma_p$ is the angle $(\mathbf{r}_{pq},\mathbf{r}_{pr})$,  $\mathbf{r}_{pq}=\mathbf{r}_{q}-\mathbf{r}_{p}$, and so forth, and 
$B=S^2\frac{4K}{\pi}(\alpha_1^2+\alpha_2^2)$~\cite{Dommersnes99,Kim99}. When wished, $U_3^{\rm pow}$ is added to the two-body contibutions of equation~(\ref{repulse}). In this work, we only consider 
two- and three-body forces because they dominate at large separations. The more general case of many-body forces will be addressed in the Discussion.

In order to take protein diversity into account, we also consider in section~\ref{short} the case were several protein species are present in the membrane. Forces then depend on the nature of the interacting proteins. On the one hand, we can play on the long-range terms by making $\varepsilon_r$ in~(\ref{biexp}) or $C$ in~(\ref{repulse}) depend on interacting protein species, through their contact angles $\alpha$. On the other hand, we can modulate the short-range attraction by using species-dependent values of  $\varepsilon_a$ and $\varepsilon_r$ to modulate $E_{\rm min}$ at fixed $\gamma_a$, $\gamma_r$ and $E_{\rm max}$.

Note that so far, we have used the terminology ``potentials or forces at long range''. However, our potentials are not long-ranged in the strict statistical mechanics sense of the word because, even though their range extends beyond several tens of nanometers, the integral $\int rU(r) {\rm d}r$ is convergent at large $r$. However, we shall keep the terminology ``long-range'' in the following, by opposition to ``short-range".

We carry out Monte Carlo simulations in the canonical ensemble as prescribed in reference~\cite{PRE08A}. More details can be found in the Supporting Information. Our systems contain $N \geq 100$ particles that interact via the potentials discussed above, and evolve in a two-dimenional continuous medium of area $\mathcal{A}$ representing the lipidic ``sea'' (the solvent). Boundary conditions are periodic. The hard-core diameter is chosen as
$d_0 = 2a=4$~nm, the typical diameter of a protein of average molecular
weight~\cite{Alberts}. Figures~\ref{exs0}, \ref{snaps}, \ref{snaps2coul} and S2-3 display several simulation snapshots for various parameter sets. The distributions of cluster sizes $P(k)$ displayed below are defined as 
\begin{equation}
P(k)=\langle N_k \rangle \bigg/ \sum_{k=1}^\infty \langle N_k \rangle,
\label{defp:k}
\end{equation}
where $\langle N_k \rangle$ is the measured average number of clusters of size $k$. The mean cluster size is defined as   $\langle k \rangle = \sum k P(k)$. It is also equal to the number of particles, $N$, divided by the number of clusters (including monomers).

\section{Long-range potentials}
\label{intermediate}

In this section, we focus on the role of long-range repulsive potentials on cluster phases. As motivated above, we first compare two kinds of pairwise potentials: exponentially [equation~(\ref{biexp})] and algebraically [equation~(\ref{repulse})] decaying ones. In a second time, we study the role of three-body forces at long range [equation~(\ref{3b})]. Finally, we focus on the implication of protein diversity as far as long-range forces are concerned. More precisely, we take into account the fact that a macroscopic fraction of up-down symmetric proteins might not be concerned by the elastic long-range repulsion.

\subsection{Comparison between two kinds of long-range pair potentials}

We first compare exponentially decaying forces and algebraically decaying ones ($U(r) \sim r^{-4})$, associated with the repulsion due to the elastic deformations of the membrane. The $N$ proteins are identical. Bi-exponential potentials have already been numerically studied in detail in references~\cite{Sear99,PRE08A,Imperio07}. The main feature of the ensuing cluster phases is that cluster-size distributions $P(k)$ appear to be bimodal for broad ranges of parameters and concentrations: large, dense clusters co-exist with a gas of monomers. Figure~\ref{Hists:N:M} below will provide related distribution examples. In addition,  let us denote by $\phi\equiv Nd_0^2/\mathcal{A}$ the (reduced) protein density. Above a limiting value $\phi^c$, a nice approximate proportionality regime, $\langle k \rangle \simeq \phi/\phi^c$, can in general be observed on a few decades in this bi-exponential case. These observations have been given theroretical interpretations in references~\cite{PRE08B,Archer08}. 

\begin{figure}[h]
\begin{center}
\includegraphics[width=0.4\textwidth]{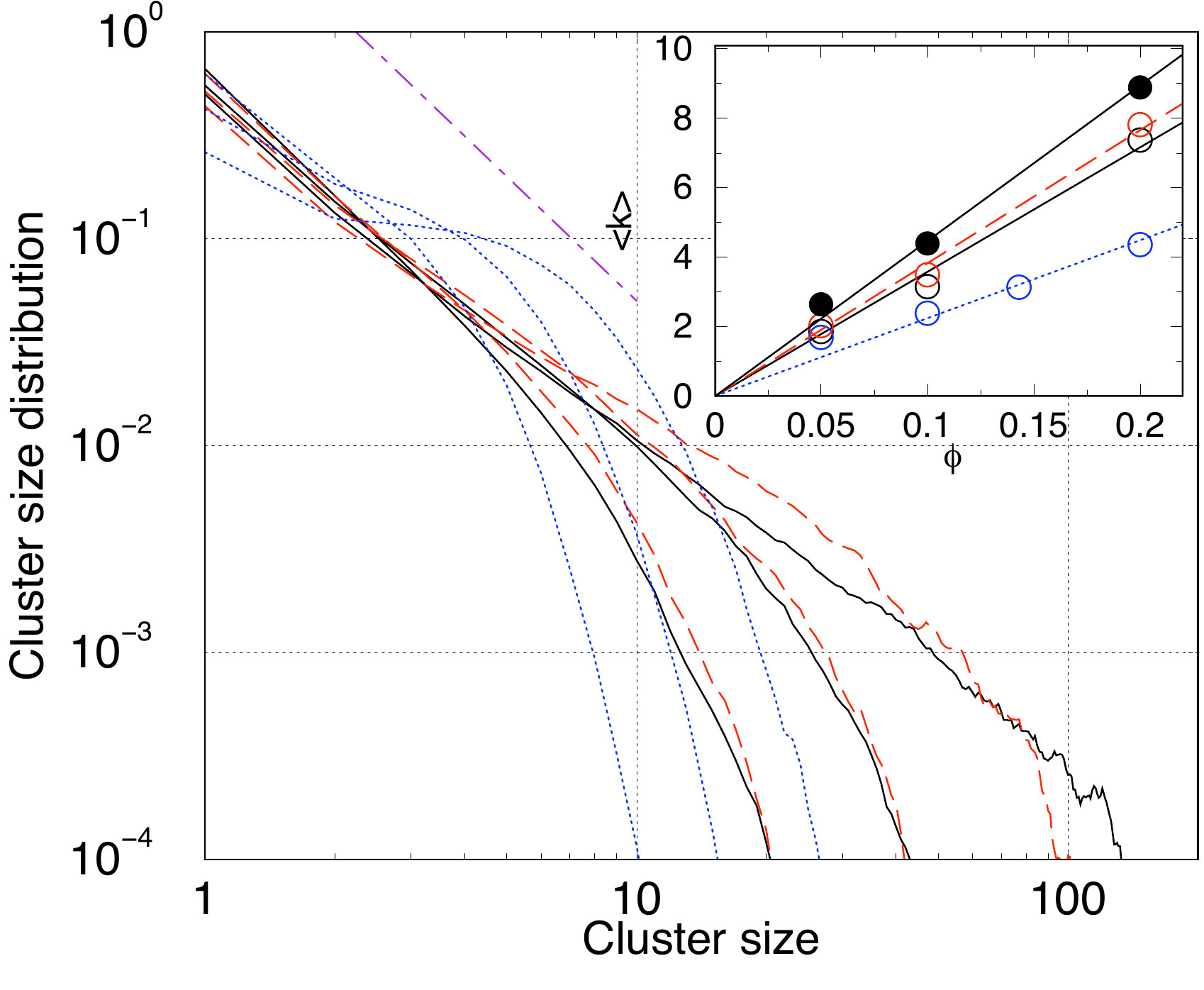} 
\includegraphics[width=0.4\textwidth]{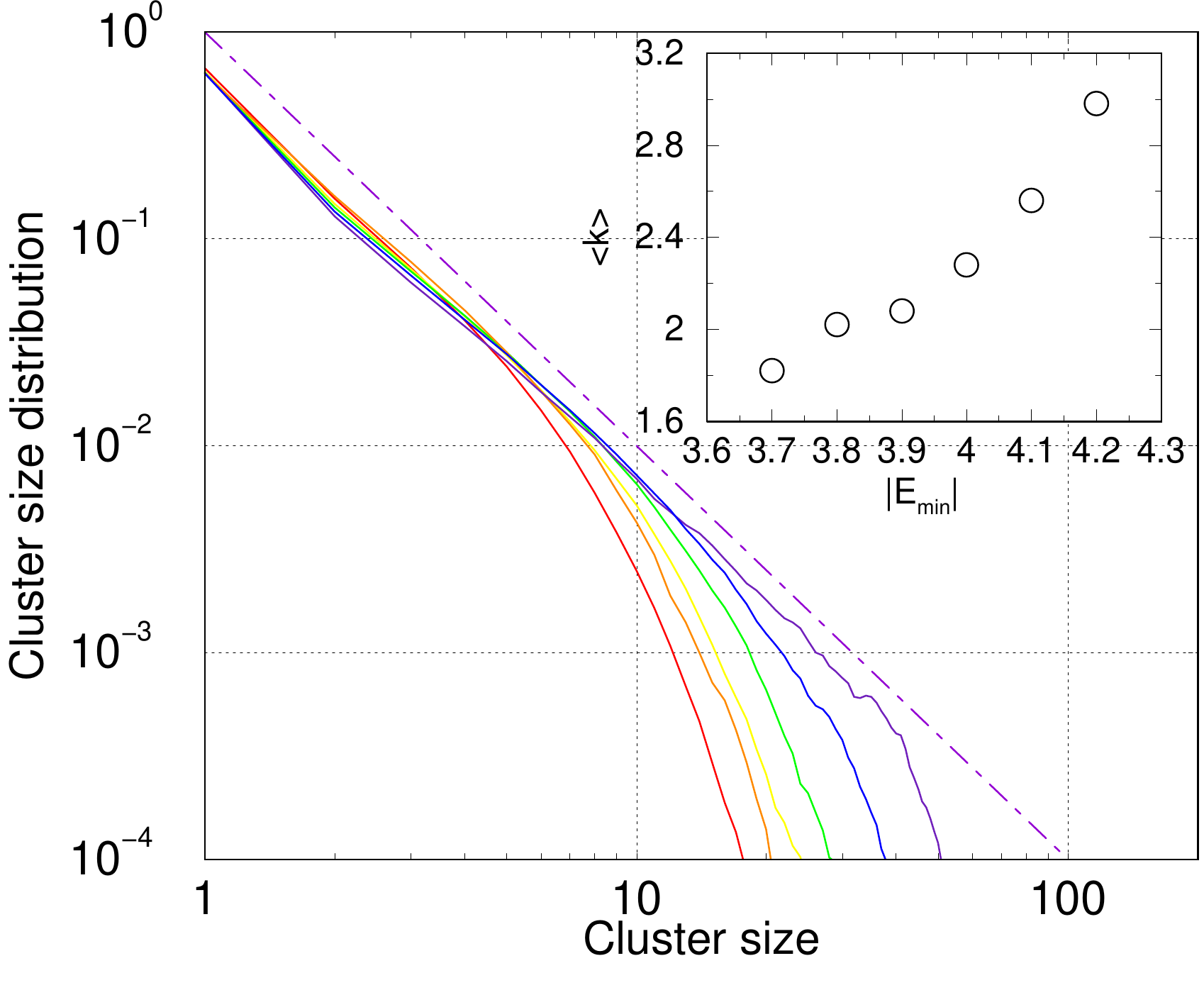} 
\caption{Distributions of cluster sizes $P(k)$ in the case of algebraic long-range repulsion. The dot-dashed lines have slope -2, for comparison. Top: for three different sets of parameter values: $K=30 \kB T$, $E_{\rm min}=-3.5 \kB T$ and thus $E_{\rm max}=0.21 \kB T$ (full black lines); $K=50 \kB T$, $E_{\rm min}=-3.8 \kB T$ and $E_{\rm max}=0.40 \kB T$ (dashed red lines); and  $K=150 \kB T$, $E_{\rm min}=-4 \kB T$ and $E_{\rm max}=1.34 \kB T$ (dotted blue lines). And $1/\gamma_a =2$~nm in all cases. From left to right at high cluster size, for each set of parameters, $\phi = 0.05$ ($N=100$), 
$\phi = 0.1$ ($N=100$) and $\phi = 0.2$ ($N=200$ except for $K=150 \kB T$ where $N=100$). Log-log coordinates. Inset: Average cluster size $\langle k \rangle$ as a function of the density $\phi$. Same color and line styles as above. For comparison, we have also plotted in this inset (filled black circles) $\langle k \rangle$ for the bi-exponential potential (blue curve) of figure~\ref{exs0}. Bottom: $K=50 \kB T$, $N=100$ and $\phi=0.05$; From left to right at high cluster size: $E_{\rm min}=-3.7$, $-3.8$, $-3.9$, $-4$, $-4.1$ and $-4.2\kB T$. Inset:  Average cluster size $\langle k \rangle$ as a function of  $E_{\rm min}$.  
\label{dist_1surR4A}}
\end{center}
\end{figure}

For particles experiencing an algebraic repulsion at long range,  the situation is somewhat different. Clusters still appear to be stable at equilibrium, since they exist independently of the initial configuration: an initial gas partially condenses into a cluster phase; converesly, a unique initial condensed droplet splits up into smaller clusters (see Supporting Information, figure S3). And when algebraic and bi-exponential potentials are globally comparable, as the red and blue potentials in figure~\ref{exs0}, systems also look similar, as illustrated by the snapshots in the same figure (see also the insets of figure~\ref{dist_1surR4A} where values of $\langle k \rangle$ are plotted). 

However, cluster-size distributions $P(k)$ are not bimodal any longer in the algebraic case, but they rather display a broad power-law-like regime with an exponent close to $-2$, until a maximum size, as illustrated in figure~\ref{dist_1surR4A}
(we have nevertheless observed that a bimodal regime is restored for large bending moduli $K$ and large $\phi$ (see figure~\ref{dist_1surR4})).  Thus can we speak of ``cluster phases"? Defining unambiguously a cluster phase is not an easy task because it is not even clear that such phases are characterized by a true thermodynamic transition at $\phi^c$~\cite{EPL10,Archer08}. And density fluctuations in gases can also lead to transient small multimers and to a value of $\langle k \rangle$ that can be slightly larger than 1. However, upon some approximations, it has been shown that density fluctuations in a gas phase, lead to a rapidly, exponentially decaying distribution~\cite{EPL10}. In the present case, we rather observe a power-law-like regime instead, and maybe more importantly, we infer from the different data sets used in figure~\ref{dist_1surR4A} that more than 25\% (resp. 75\%) of the proteins dwell in clusters containing more than $k=10$ particles as soon as $\langle k \rangle \geq 3$ (resp. $\langle k \rangle \geq 7$). Independently of a precise definition of a cluster phase as in~\cite{EPL10,Archer08}, this indicates that a macroscopic fraction of proteins live in assemblies, which is the biological mechanism we are primarily interested in.

\begin{figure}[h]
\begin{center}
\includegraphics[width=0.5\textwidth]{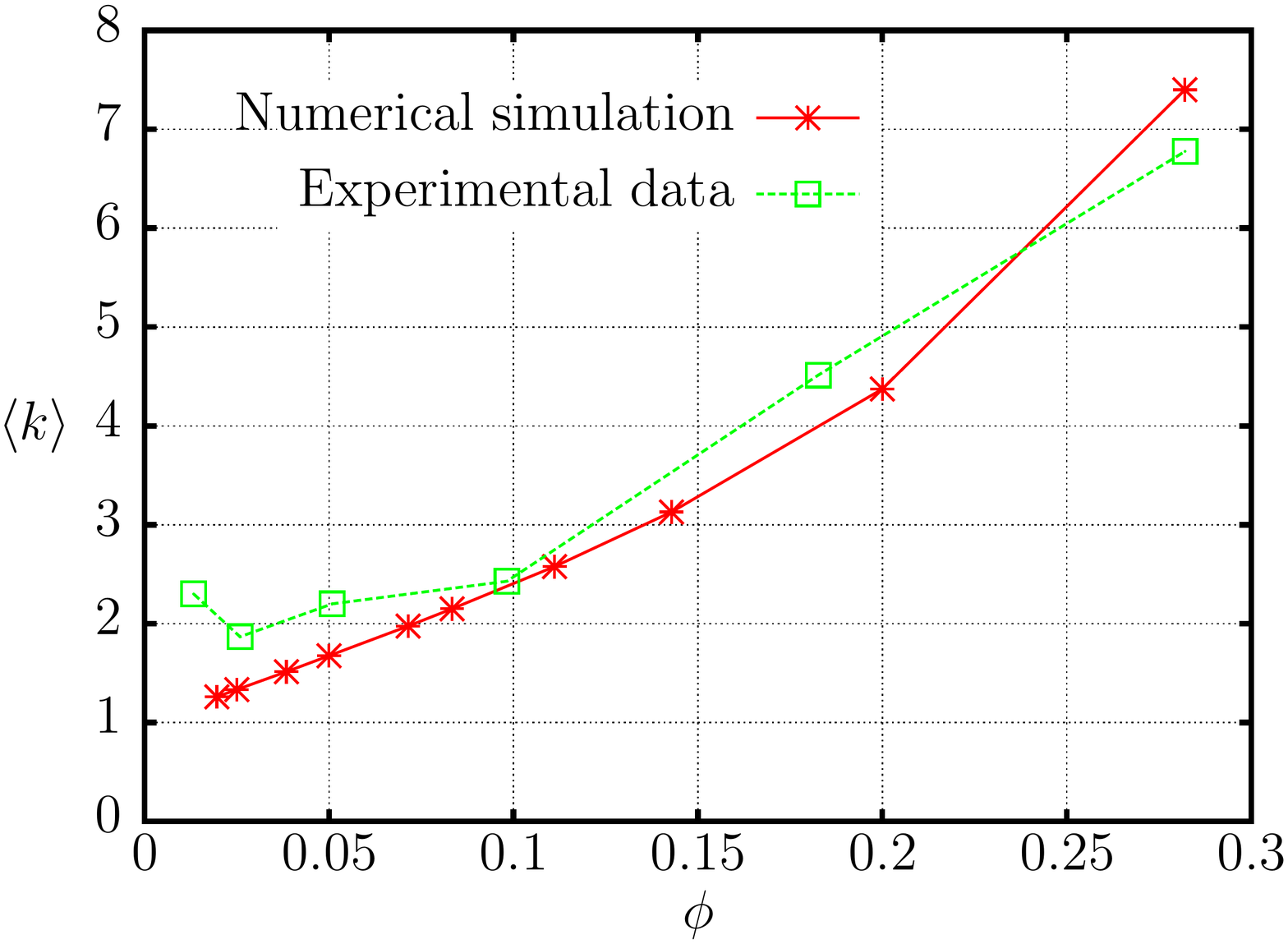} \\ \vspace{-1cm}
\caption{Mean cluster size $\langle k \rangle$ as a function of the density $\phi$. Comparison bewteen simulation data for algebraic long-range repulsion, with $K=150 \kB T$, $1/\gamma_a=2$~nm, $E_{\rm min}=-4 \kB T$, and $N=100$, and experimental measurements, inferred from~\cite{Gulik87} (see text). 
\label{Gulik}}
\end{center}
\end{figure}

The proportionality regime is still being observed, even though on shorter concentration ranges (see figure~\ref{dist_1surR4A}, top, inset, and figure~\ref{Gulik}). At the highest density studied, $\phi=0.2$, average cluster sizes $\langle k \rangle$ range from 4 to 8 for the parameter sets studied and clusters of size up to $k=20$ are commonly observed. Note that lowering $E_{\rm min}$ at fixed $K$ (i.e. increasing the attraction) should increase $\langle k \rangle$~\cite{PRE08B}. Unfortunately, due to the long-tail distribution, increasing $\langle k \rangle$ requires increasing $N$, which is rapidly limiting in terms of computational cost. However, we have studied the effect of varying  $|E_{\rm min}|$ in the $K=50 \kB T$ and $\phi=0.05$ case (figure~\ref{dist_1surR4A}, bottom). The power-law-like regime extends as  $|E_{\rm min}|$ grows and $\langle k \rangle$ increases concomitantly (see Inset).

To conclude, this numerical study indicates that replacing and exponentially decaying repulsion at long range by an algebraic one does not alter the existence of cluster phases for wide ranges of parameters, thus demonstrating that they should be robust with respect to the precise shape of interaction potential, provided that it displays both a short-range ($\sim~1$~nm) attraction and a longer-range ($\sim~10$~nm), weaker repulsion.

To finish with, we compare our data to available experimental ones. Clusters of bacteriorhodopsin (BR) in proteoliposomes can be observed by freeze-fracture electron microscopy. This technique makes possible the determination of cluster-sizes and of their distributions with a correct accuracy~\cite{Gulik87} (see Table I and figure~4 in this reference). We assume a BR radius of 1.7~nm~\cite{Gulik87} to convert numbers of BR per $\mu$m$^2$ to densities $\phi$. The algebraically decaying potential~(\ref{repulse}) has been chosen for simulations, with $K=150 \kB T$, $1/\gamma_a=2$~nm and $E_{\rm min}=-4 \kB T$. With these parameter values, a good qualitative agreement is obtained, as displayed in figure~\ref{Gulik}. Cluster-size distributions are bimodal in both cases, with a monomer peak and clusters containing a few proteins (compare both our green curve in figure~\ref{dist_1surR4} and figure~4 of~\cite{Gulik87}). This good qualitative agreement supports our numerical approach. Similar cluster sizes ($k \approx 10$) have been observed at lower protein density but after photoactivation of BR~\cite{Kahya02}.

\subsection{Role of three-body elastic interactions}

It has been proved~\cite{Dommersnes99,Kim99} that the elastic energy for isotropic inclusions is dominated at long range by the sum of the two-body terms of equation~(\ref{repulse}) considered so far and the three-body ones of equation~(\ref{3b}). Here we take these three-body terms into account in simulations. Contrary to two-body forces, we did not include any cut-off at short distances, the hard-core repulsion playing this role. Note that in this paragraph, $E_{\rm min}$ still refers to the two-body energy at contact of $U_2^{\rm pow}(2a)$ [equation~(\ref{repulse})].

Simulating three-body forces consists of an important increase of the computational complexity, which grows from $\mathcal{O}(N^2)$ to $\mathcal{O}(N^3)$ operations per Monte Carlo step. For this reason, systems with $N > 100$ can hardly be simulated and we could perform a few simulations only, in order to check that cluster phases still exist in this case and thus that two-body repulsion dominates three-body forces, which can be attractive or repulsive (see also the Supporting Information). 

\begin{figure}[ht]
\begin{center}
\includegraphics[width=0.4\textwidth]{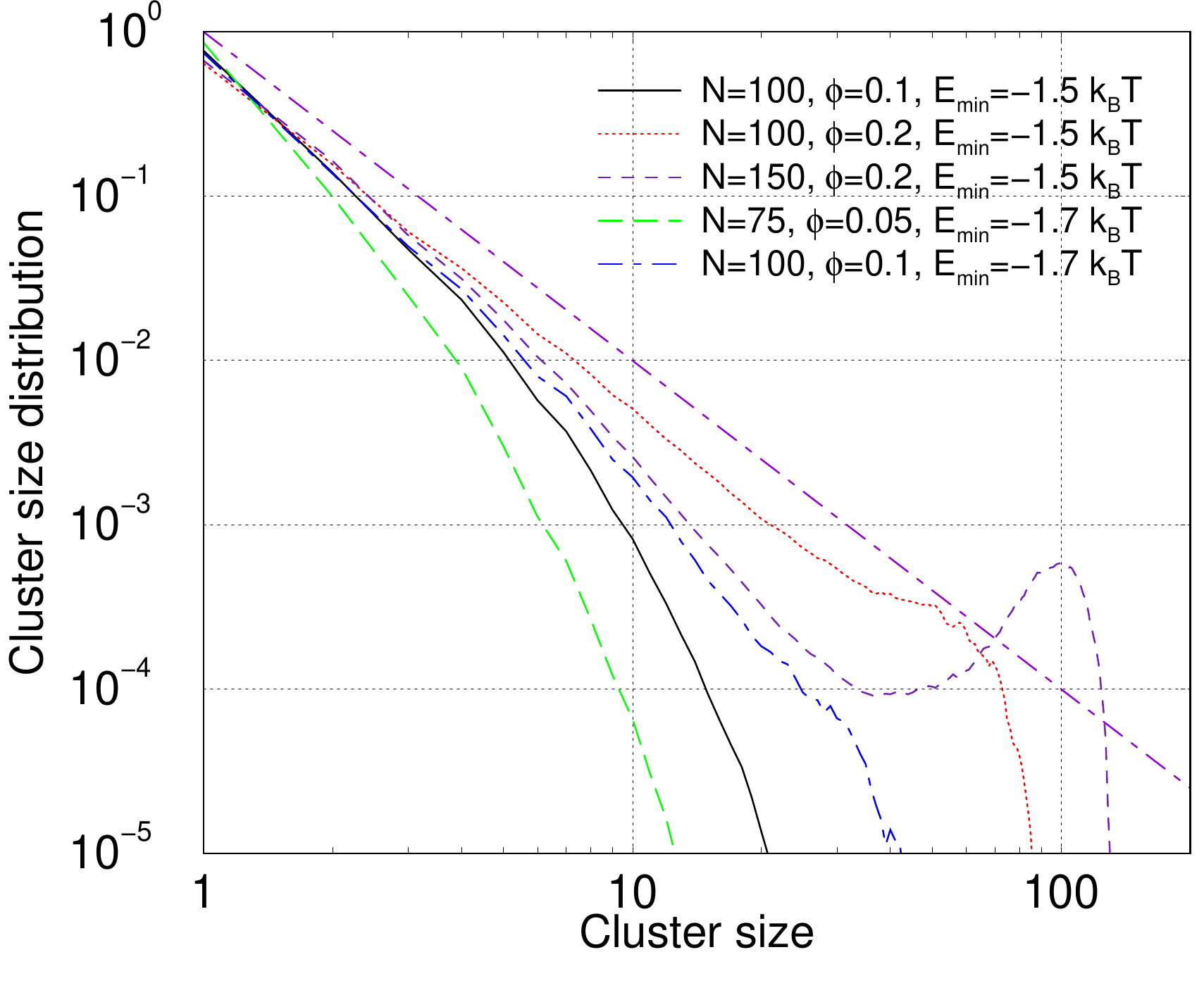} 
\caption{Distributions of cluster sizes $P(k)$  in the case of algebraic long-range interaction including three-body interaction terms of equation~(\ref{3b}), for different parameter sets. In all cases, $K=30 \kB T$. The dot-dashed line has slope -2 for comparison. Log-log coordinates. 
\label{dist_1surR4_3c}}
\end{center}
\end{figure}

The principal features observed in the previous subsection are conserved. Cluster phases appear to be stable at equilibrium and their distributions display a power-law-like regime until a maximum size. Examples are provided in figure~\ref{dist_1surR4_3c}. When increasing the density, the mean cluster-size $\langle k \rangle$ grows, even if we do not have enough points at our disposal to conclude that a proportionality regime still exists.  Lowering $E_{\rm min}$ at fixed $K$ also increases $\langle k \rangle$. 

However, some differences are also noticeable. First, higher values of the energy at contact $E_{\rm min}$ are required to obtain distributions comparable to those of the previous paragraph. Their are close to $-1.5 \kB T$ instead of $-4 \kB T$. This indicates that three-body forces have an attractive contribution, on average, as already noticed in reference~\cite{Kim98} and in the Supporting Information. In addition, at high $\phi$, simulations show that a big cluster, containing a large fraction of the $N$ particles, can transiently appear and subsequently disintegrate into smaller clusters. As a consequence, one distribution displays a secondary peak at large cluster sizes in the figure, for $\phi=0.2$, $E_{\rm min}=-1.5 \kB T$ and $N=150$ (whereas it is absent if $N=100$). This might be the signature of a transition between a cluster phase and a liquid-gas coexistence phase when increasing $\phi$ \cite{Archer08}. To clarify this point, simulating larger systems will be required. But again, simulating three-body forces is expensive and our statistics are poorer than in the previous sections. 

%% J'AI FAIT 2.5 10^7 MCSweeps

\subsection{Modulation of long-range potentials}

\begin{figure}[ht]
\begin{center}
\includegraphics[width=0.4\textwidth]{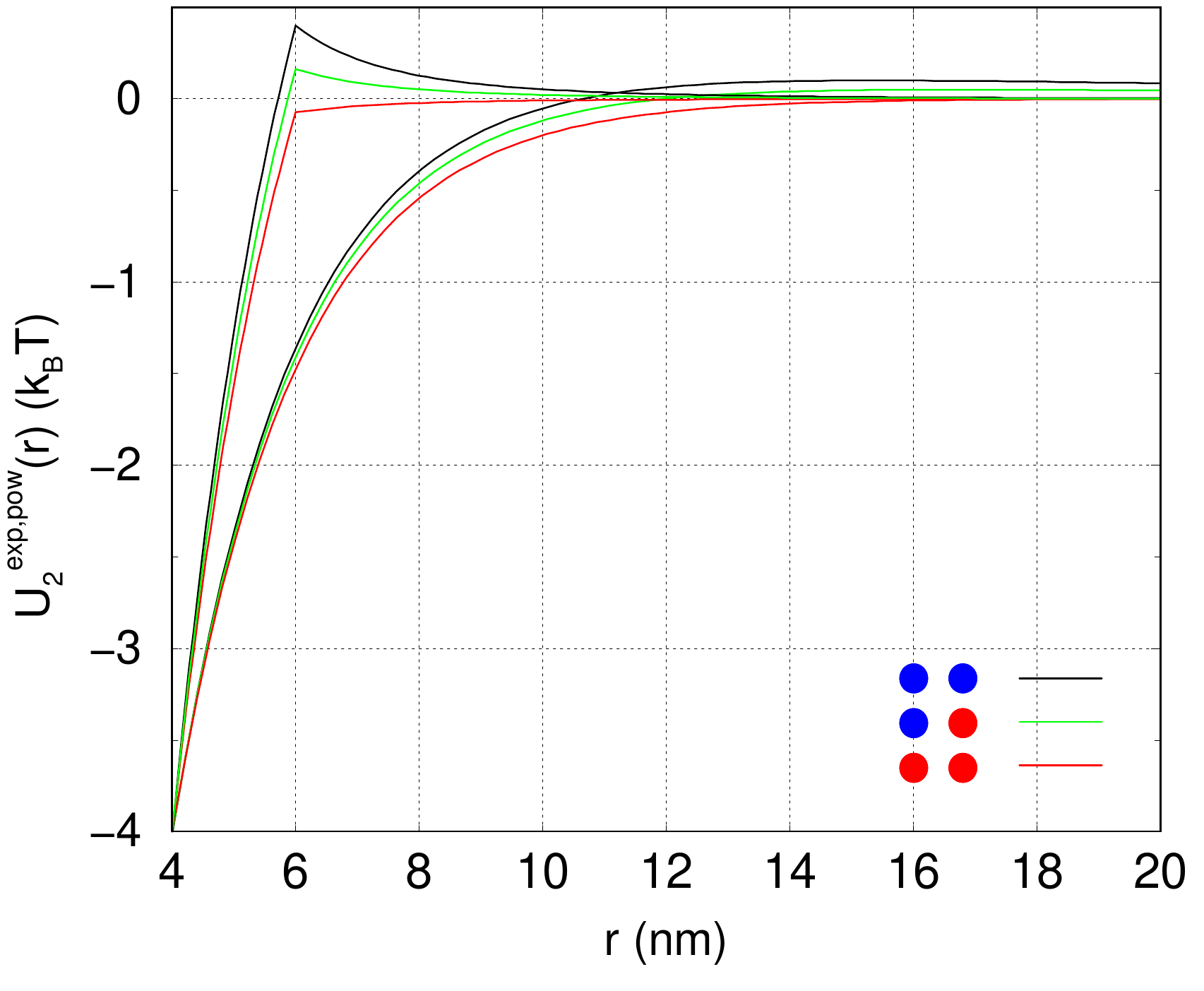}   
\caption{Algebraic pair potentials $U_2^{\rm pow}$~(three upper curves) and bi-exponential ones $U_2^{\rm exp}$~(lower curves) modulated at long range. This modulation takes into account the fact that cylindrical inclusions (in red) are up-down symmetric ($\alpha_R=0$) while conical ones (in blue) are asymmetric ($\alpha_B=10^{\circ}$), as in the previous paragraphs. Here $E_{\rm min}=-4 \kB T$ and $1/\gamma_a=2$~nm for all curves, $K=50 \kB T$ for the algebraic repulsion, $E_{\rm max}=0.1 \kB T$, $0.05 \kB T$ and 0 for the three bi-exponential potentials, respectively (and $1/\gamma_r=16$~nm).
\label{modul:long}}
\end{center}
\end{figure}

We now explore the consequences of protein diversity in terms of long-range forces, for both kinds of pair potentials (bi-exponential, equation~(\ref{biexp}), and algebraic, equation~(\ref{repulse})). Examples of modulated potentials are displayed in figure~\ref{modul:long}. 

In the algebraic case, the modulation is ensured by taking $\alpha=0$ (cylindrical inclusions thereafter) instead of $\alpha=10^\circ$ (conical inclusions) when calculating the constant $C$. This constant is essentially divided by 2 for large $K$ values when a conical and a cylindrical inclusion interact. It becomes negative for two cylindrical inclusions, which feel a mutual attraction because of Casimir forces. As for bi-exponential potentials, we adapt both $\varepsilon_a$ and $\varepsilon_r$ so that $E_{\rm min}$ remains unchanged whereas $E_{\rm max}$ is divided by 2 for cylindrical-conical interactions and $E_{\rm max}=0$ for cylindrical-cylindrical ones. In the latter case, we simply set $\varepsilon_r=0$ and adapt $\varepsilon_a$ to set the value of $E_{\rm min}$; The interaction is everywhere attractive. We denote by $\NR$ and $\NB$ the numbers of cylindrical and conical particles and $N=\NR+\NB$. We compare systems with a finite cylindrical-inclusion fraction $\xR \equiv \NR/N$ to systems of $N$ identical conical particles, as above.

\begin{figure}[h!]
\begin{center}
\vspace{-1.3cm}
\includegraphics[height=8.2cm,width=8.2cm]{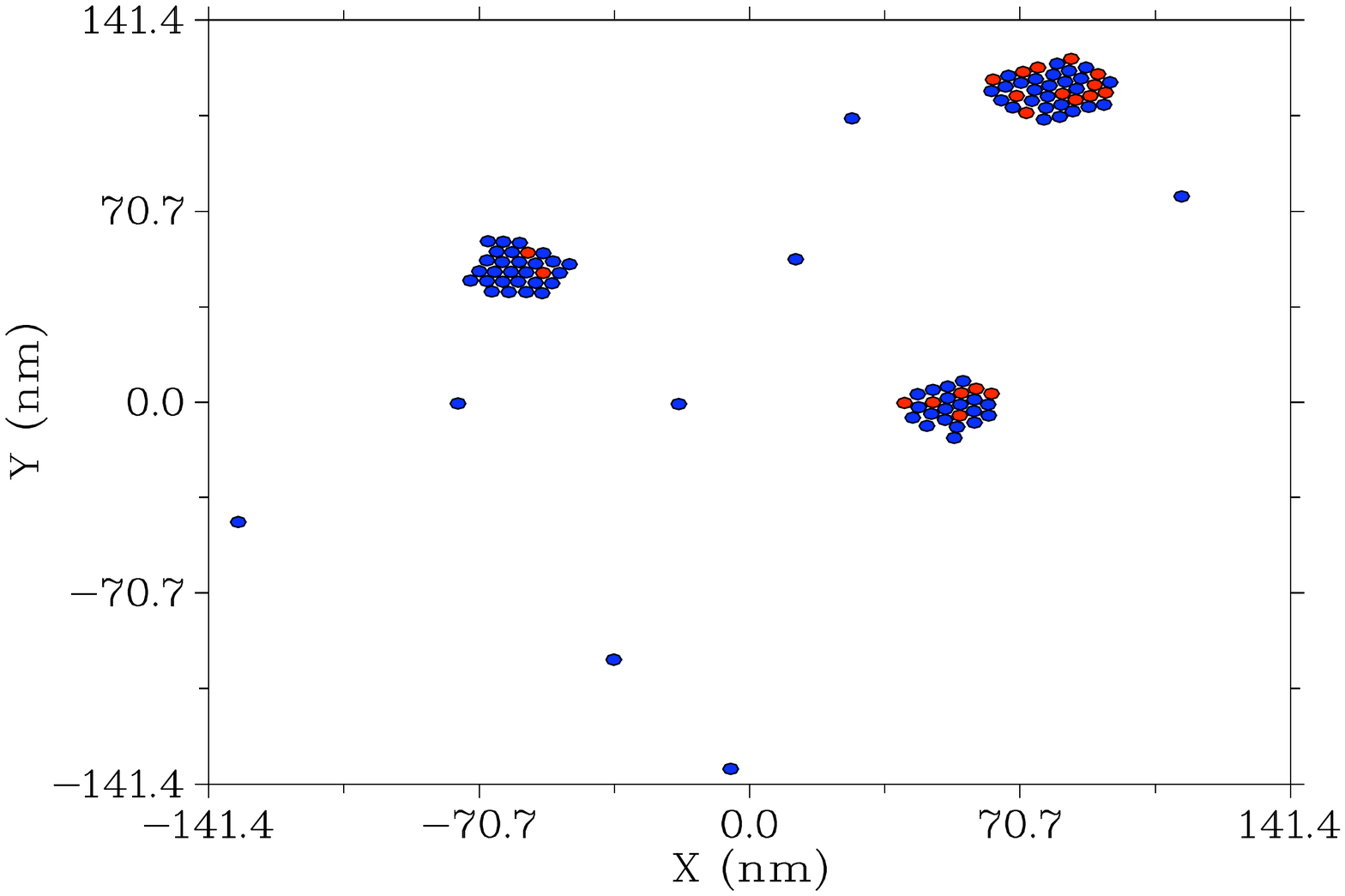}   \\ \vspace{-2cm}
\includegraphics[height=8.2cm,width=8.2cm]{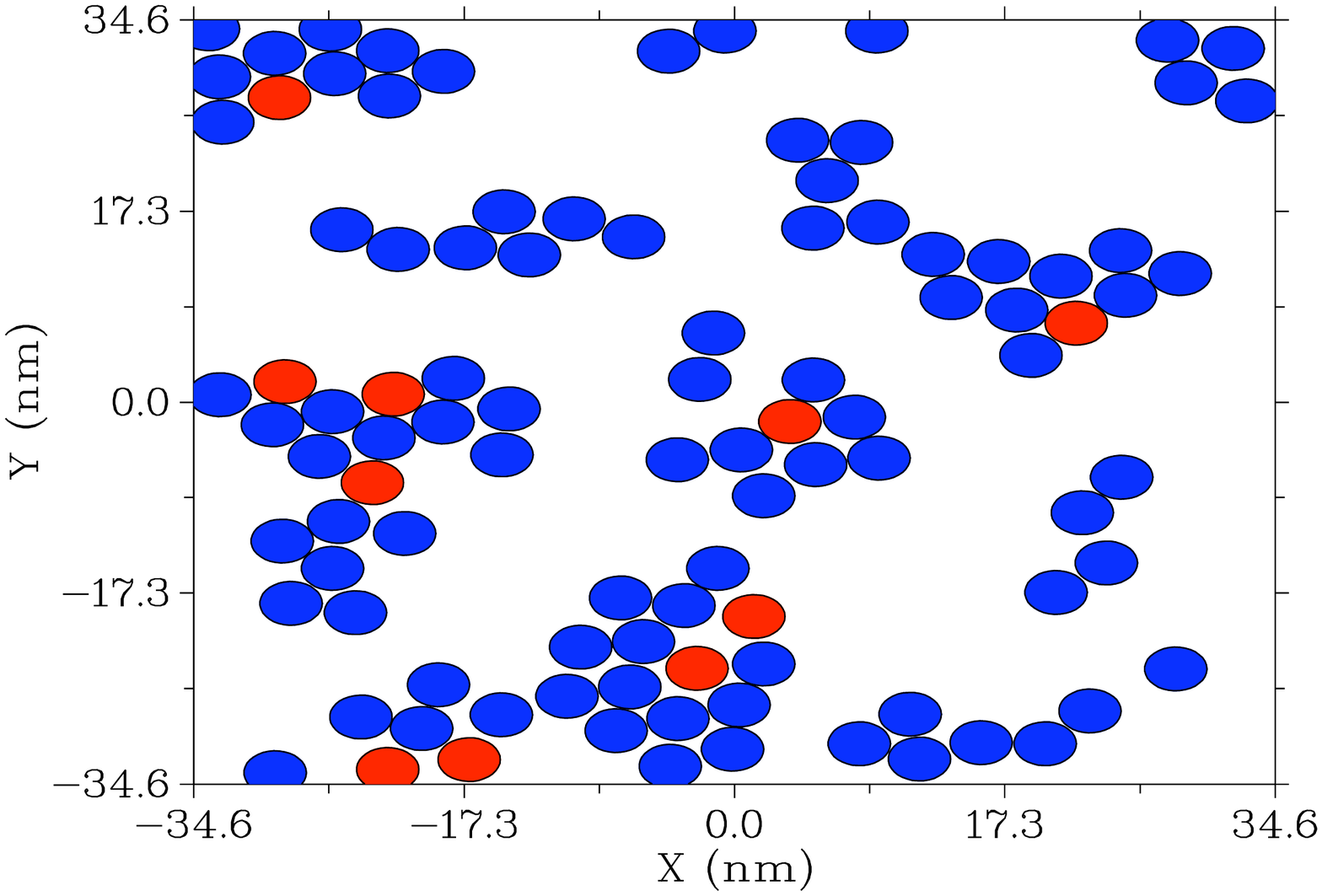} \\ \vspace{-2cm}
\includegraphics[height=8.2cm,width=8.2cm]{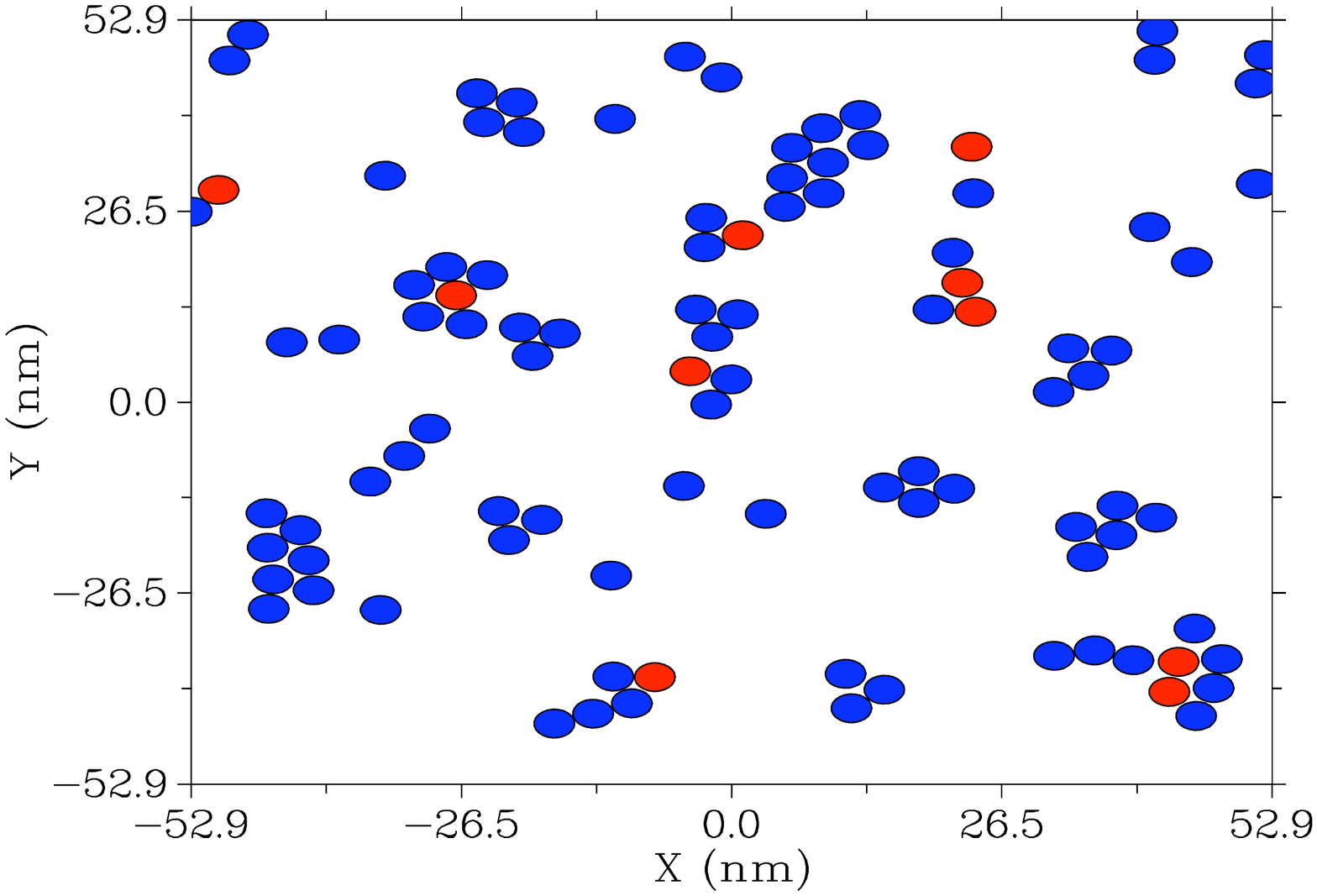} \\ \vspace{-1cm}
\caption{Top: Snapshot of a cluster phase with $\NR=20$ cylindrical inclusions (in red) and $\NB=80$ conical ones (in blue), in the case of bi-exponential pair potentials as in figure~\ref{modul:long}. The density is $\phi=1/50$. Middle and bottom: Snapshots of cluster phases with $\NR=10$ cylindrical inclusions and $\NB=90$ conical ones, in the case of the pairwise algebraic repulsion with $K=150 \kB T$, $E_{\rm min}=-4 \kB T$ and $1/\gamma_a=2$~nm. Densities are $\phi=1/3$ (middle) and $\phi=1/7$ (bottom). In all three cases, mono-particle systems ($\NR=0$) give essentially similar snapshots.
\label{snaps}}
\end{center}
\end{figure}

\begin{figure}[h!]
\begin{center}
\vspace{-0.2cm}
\includegraphics[width=0.35\textwidth,angle=-90]{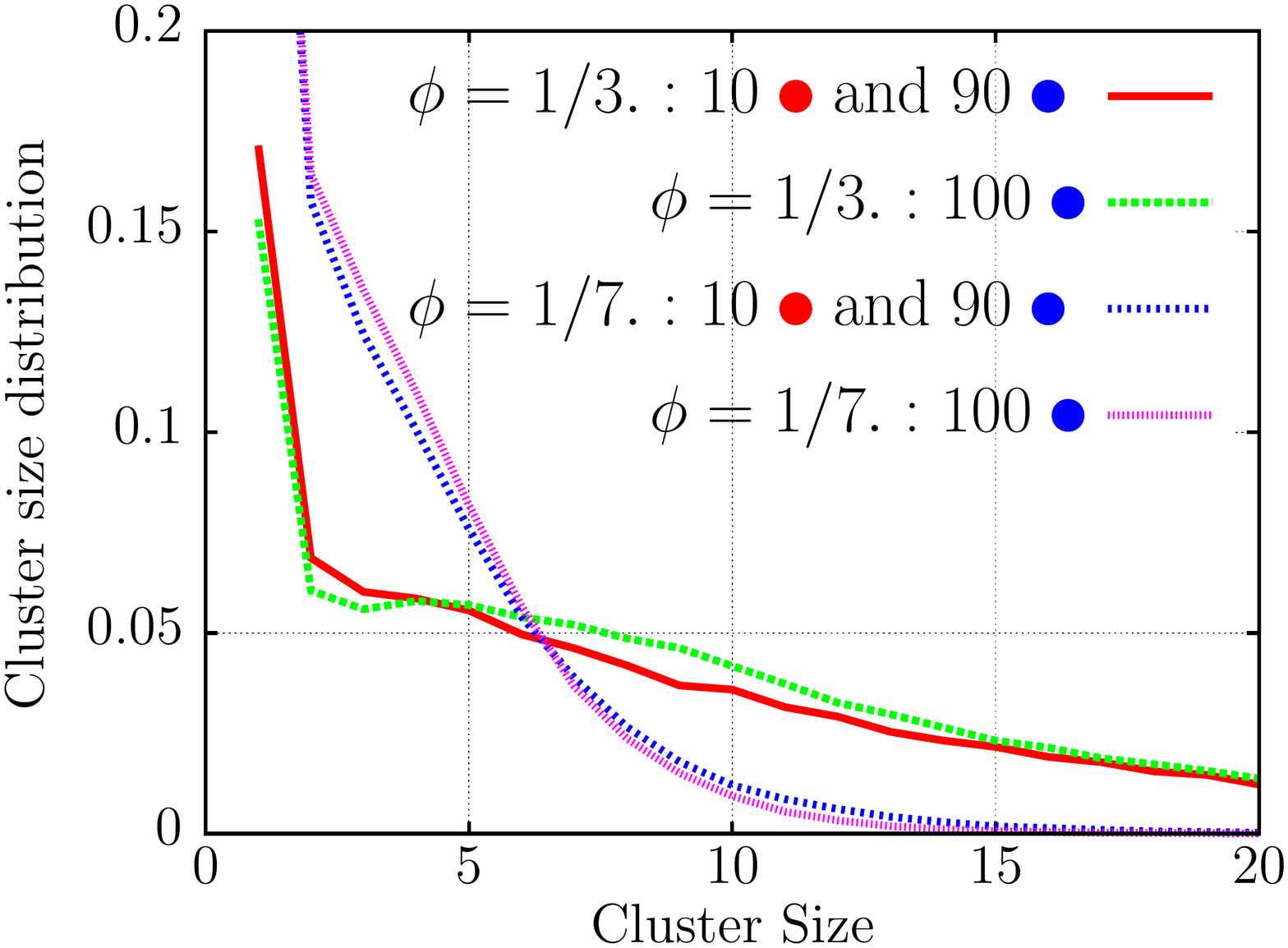} \\Ê\vspace{-7mm}
\caption{Distributions of cluster sizes $P(k)$ in the case of the pairwise algebraic long-range repulsion with $K=150 \kB T$, $E_{\rm min}=-4 \kB T$ and $1/\gamma_a=2$~nm, for two different densities $\phi$ and two different cylindrical inclusion numbers $\NR$ ($N=100$ particles in all cases). The green distribution is (weakly) bimodal.
\label{dist_1surR4}}
\end{center}
\end{figure}

\begin{figure}[ht]
\begin{center}
\vspace{-0.2cm}
\includegraphics[width=0.5\textwidth]{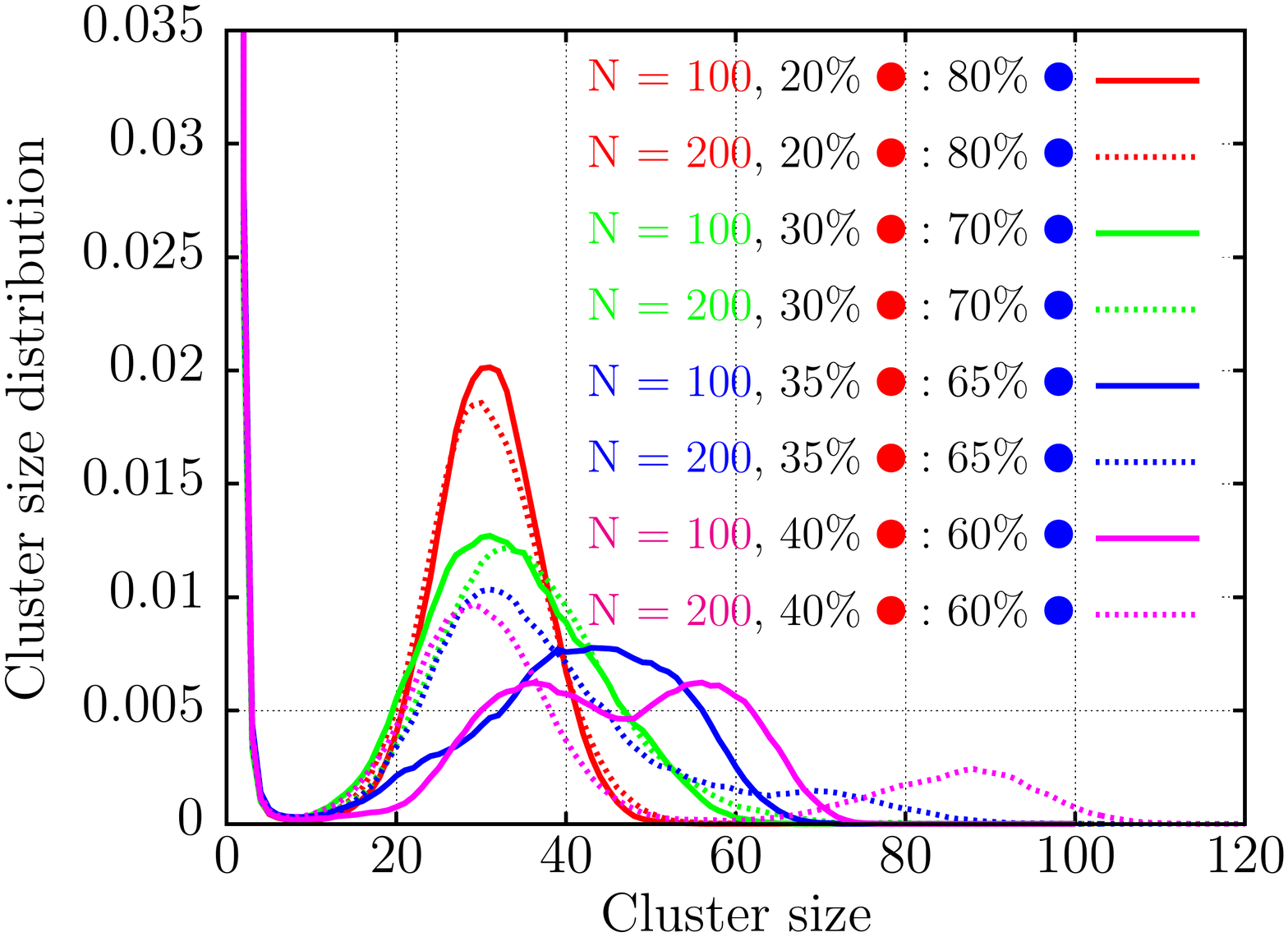} \\ \vspace{-0.4cm}
\includegraphics[width=0.5\textwidth]{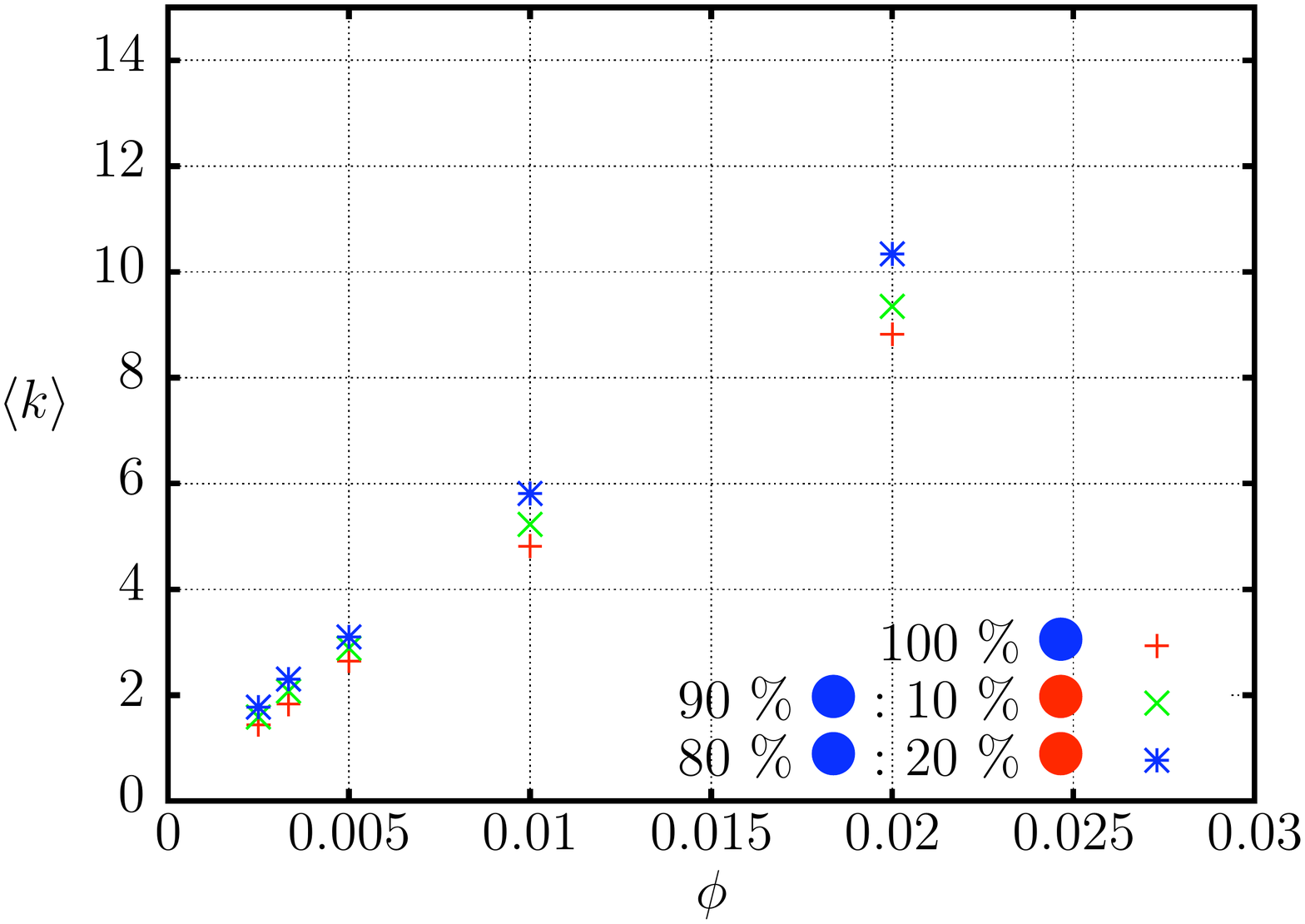} \\ \vspace{-1cm}
\caption{Top: Distributions of cluster sizes in the case of bi-exponential pair interaction potentials as in figure~\ref{modul:long}, with density $\phi=1/50$, and various cylindrical inclusion fractions $\xR$. Comparison between $N=100$ (soild lines) and $N=200$ (dotted lines) for each value of $\xR$. Bottom: resulting mean cluster-size as a function of the density, exhibiting a proportionality regime. 
\label{Hists:N:M}}
\end{center}
\end{figure}

Figure~\ref{snaps} provides simulation snapshots. For small cylindrical-inclusion fractions $\xR$, they are very similar to the mono-particle case as studied above. Figure~\ref{dist_1surR4}  shows examples of cluster-size distributions $P(k)$ that are weakly affected by the presence of a small amount of cylindrical inclusions, at fixed $\phi$.  Consequently, proportionality regimes are also preserved (figure~\ref{Hists:N:M}, bottom).

To interpret these numerical observations, we discuss them in the framework of the liquid-gas transition of binary mixtures, even though in the present context of cluster phases, transitions are certainly not true thermodynamic transitions but crossovers~\cite{EPL10}. We see a cluster as a liquid droplet co-existing with gas. In the liquid phase, conical and cylindrical inclusions are miscible because they essentially feel the same short-range attraction. This miscibility is visible on simulation snapshots (figure~\ref{snaps}). However, the attraction is slightly stronger for cylindrical inclusions. Their second virial coefficient
$B_2=\pi \int_0^\infty r [1-\exp(-U(r))] {\rm d}r$ is thus slightly weaker, and the transition occurs at a lower density. Consequently, clusters are bigger and enriched in cylindrical inclusions, as observed. The average fraction $\xR^{\rm cl}$ of cylindrical inclusions {\em within} clusters is $> \xR$. However, conical inclusions remain the majority in clusters and too large clusters are unstable, in spite of the cylindrical-inclusion fraction. The presence of a minority of such particles does not destabilize the cluster phase.

However, we anticipate that there exists a value of $\xR$ above which this mechanism will cease being valid. To test this hypothesis, we increased the fraction of cylindrical inclusions up to $\xR=0.4$. Figure~\ref{Hists:N:M} shows our results for bi-exponential potentials, for both $N=100$ and 200 particles.  As $\xR$ grows, the cluster peak drifts to larger sizes in the bimodal distributions for $N=100$. But the $N=200$ simulations indicate that this large peak is in fact the superimposition of two peaks for $\xR>0.3$. And the inspection of simulation snapshots indicates that there are indeed two kinds of clusters: mixed clusters as in the low $\xR$ regime, the size distribution of which matches the low $\xR$ one; and one larger assembly that is highly enriched in cylindrical inclusions. This indicates the existence of a saturation phenomenon: beyond $\xR \simeq0.3$, additional cylindrical inclusions ``precipitate'' in a single large assembly with rare conical proteins. Note however that this large assembly is not stable: one observes in simulations that this cluster can disintegrate and be reformed somewhere else. This instability seems to be stronger at $\xR=0.35$ than at $\xR=0.4$. To sum up, if $\xR$ is too large, many cylindrical inclusions aggregate in a single bulk liquid phase and remaining (cylindrical and conical) inclusions form a cluster phase as above, but with a lower effective $\xR$. 

\section{Modulation of short-range potentials  and specialization of clusters}
\label{short}

We now focus on the role of short-range potential modulations in order to explore further the role of protein diversity. One of our major goals here is to validate on numerical grounds the analytical predictions of reference~\cite{EPL10} concerning protein cluster specialization: for a sufficiently large affinity difference between same- and distinct-family proteins, clusters become heterogeneous and contain essentially one protein family. We indeed consider systems with $q$ families of proteins. Each family represents proteins, not necessarily identical, that have a preferential affinity $E_{\rm min}$ at close range. There are typically $q=10^3$ families in a real membrane. The energy at contact of two particles in a same family $i$, called same-family particles hereafter, is denoted by $E_{i,i,{\rm min}}<0$. The energy at contact of particles of two distinct families $i$ and $j$ (distinct-family particles) is denoted by $E_{i,j,{\rm min}} \geq E_{i,i,{\rm min}}$.

In this section, we focus on bi-exponential potentials [equation~(\ref{biexp})] because large clusters are more abundant in this case due to their bimodal distribution. This provides better statistics when computing Binder cumulants below. We consider a single density, $\phi=1/50$, and the most symmetric case where the $q$ families contain the same number $M=N/q$ of particles. All same-family energies at contact, $E_{i,i,{\rm min}}$, are also identical, as well as distinct-family ones, $E_{i,j,{\rm min}}$. The energy barrier, $E_{\rm max}$, is unchanged and energies at contact are set by tuning $\varepsilon_a$ and $\varepsilon_r$ in~(\ref{biexp}), as explained in Section~\ref{sota}. Following~\cite{EPL10}, we introduce a Flory parameter $\chi \simeq 6(E_{i,j,{\rm min}}-E_{i,i,{\rm min}})\equiv6 \Delta E$~\footnote{This is only an approximation because this expression does not correctly take into account the entropic part of the binding free energy related to thermal fluctuations in a cluster~\cite{EPL10}.}, illustrative of the difference of affinity between same- and distinct-family proteins. In this reference~\cite{EPL10}, it was demonstrated that there exist a threshold value $\chi^c$ such that if $\chi<\chi^c$, particles of different families are mixed within clusters, whereas if $\chi>\chi^c$, they are demixed and clusters contain predominantly one family. Demixing is due to a competition between entropy of mixing and contact energy.
Note that this transition (as well as the one at $\phi^c$) is certainly a simple crossover and not a true thermodynamical transition because of the finite cluster size~\cite{EPL10}.
The theoretical prediction is $\chi^c = 2$ for $q=2$ and
\begin{equation}
\chi^c = 2 \frac{q-1}{q-2} \ln(q-1) \simeq 2 \ln q
\label{chiCtheo}
\end{equation}
for $q \geq 3$ (see also~\cite{Mittag74}). The great interest of equation~(\ref{chiCtheo}) lies in the fact that $\chi^c$ grows like $\ln k$ and $\Delta E$ remains of the order of $\kB T$ even if $q$ is large, making this scenario physically reasonable to account for cluster specialization~\cite{EPL10}. This is the reason why we wish to validate this result on numerical grounds.

\begin{figure}[ht]
\begin{center}
\vspace{-1cm}
\includegraphics[height=8.3cm,width=8.3cm]{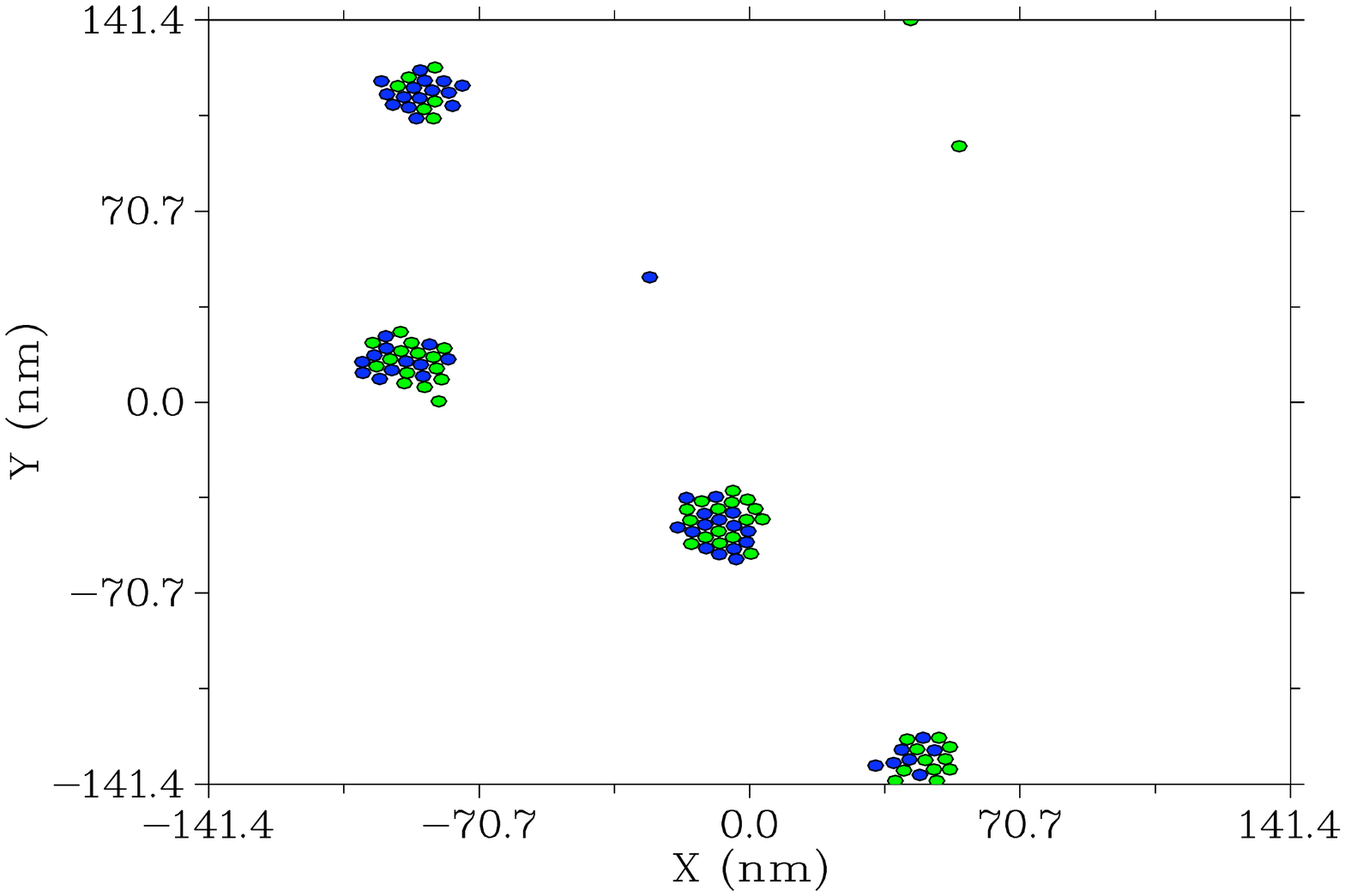}  \\ \vspace{-2cm}
\includegraphics[height=8.3cm,width=8.3cm]{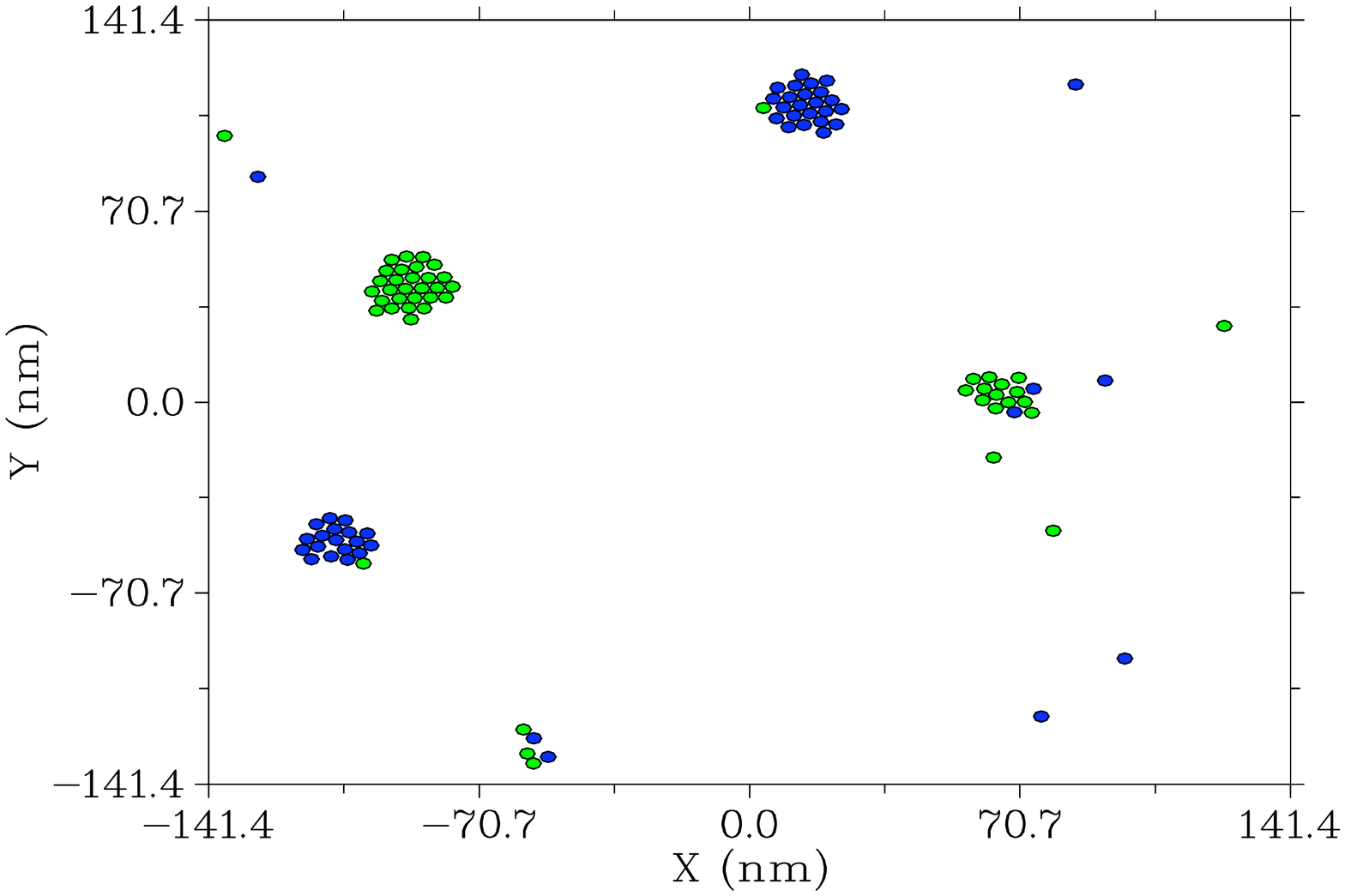} \\ \vspace{-1cm}
\caption{Snapshots of cluster phases with $q=2$ families of 50 particles each (blue and green; $N=100$), in the case of bi-exponential pair potentials with distinct-family energy at contact $E_{i,j,{\rm min}}=-3.9 \kB T$ (top, mixed clusters) and $-3.2 \kB T$ (bottom, demixed clusters). In both cases, the same-family energy at contact is $E_{i,i,{\rm min}}=-4 \kB T$, and the density is $\phi=1/50$.
\label{snaps2coul}}
\end{center}
\end{figure}

\begin{figure}[ht]
\begin{center}
\includegraphics[width=0.45\textwidth]{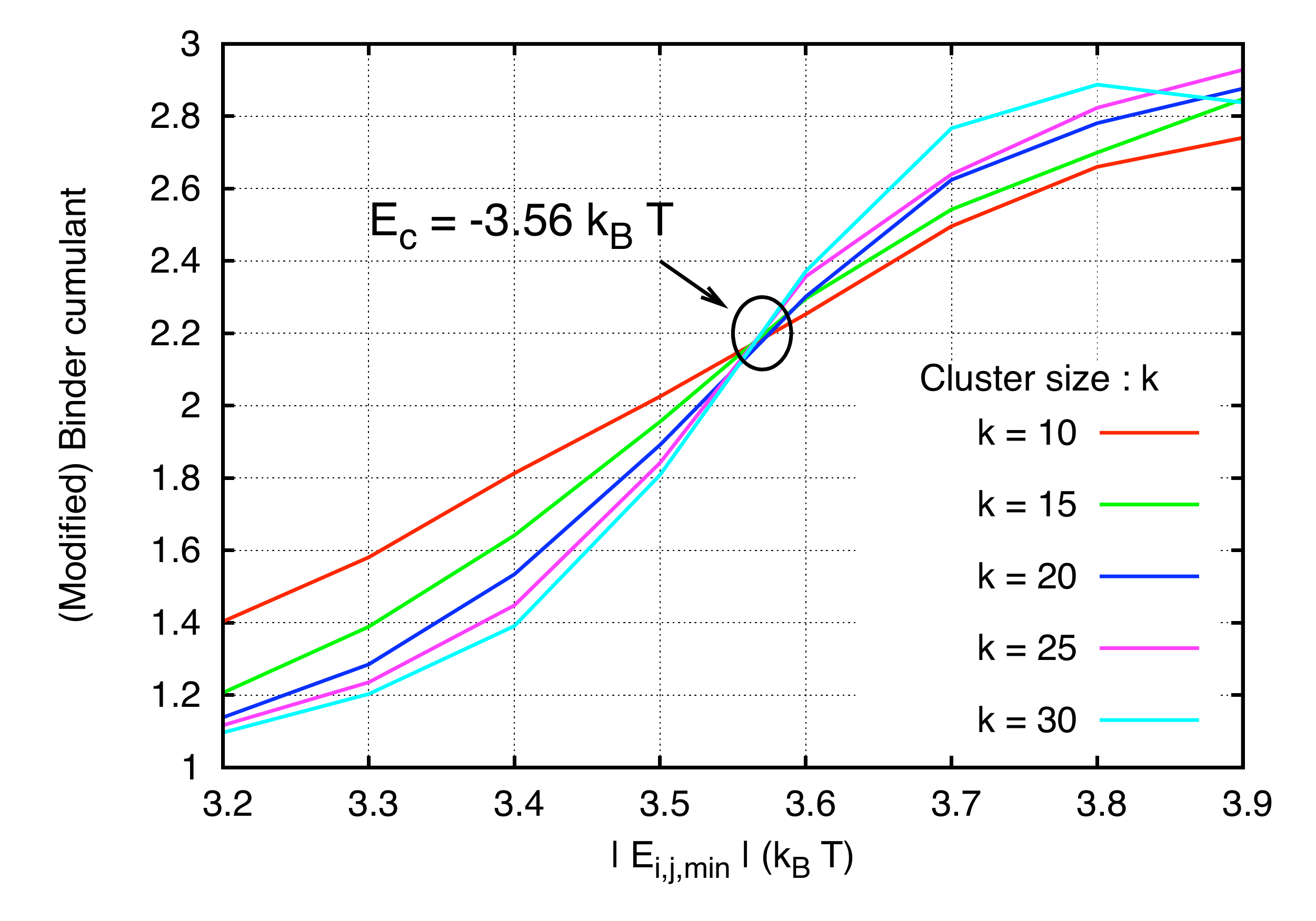} 
\caption{Modified Binder cumulants $B_k$ for $q=2$ families. Same parameters as in figure~\ref{snaps2coul}.
\label{binders2coul}}
\end{center}
\end{figure}

Figure~\ref{snaps2coul} shows two simulation snapshots in qualitative agreement with this prediction. We also simulated systems with up to $q=5$ families, showing the same agreement. To quantify further these qualitative observations, we determine the threshold value $\chi^c$ by use of (modified) Binder cumulants, an efficient tool to detect phase transitions numerically~\cite{Binder}. Our system being equivalent to a mean-field $q$-Potts model~\cite{EPL10}, we naturally use the following order parameter:
\begin{equation}
m = \left| \sum_{j=1}^q x_j \omega^{j-1} \right|, 
\end{equation}
where $x_j$ is the fraction of family $j$ proteins in a cluster and $\omega=\exp(2i\pi/q)$ is the $q$-th root of unity. The (modified) Binder cumulant is then defined as 
\begin{equation}
B_k = \frac{\langle m^4 \rangle_k}{\langle m^2 \rangle_k^2},
\end{equation}
where averages are computed on clusters of size $k$. Figure~\ref{binders2coul} displays the numerical values of $B_k$ for $q=2$ as a function of $k$ and $E_{i,j,{\rm min}}$. All curves cross at $E^c_{i,j,{\rm min}}=-3.56 \kB T$, which is the signature of a phase transition for finite-size systems (the clusters in the present case)~\cite{Binder}. The corresponding value of the Flory parameter is $\chi^c=2.64$, which compares well to the theroretical value, $\chi^c=2$. 

We followed the same procedure up to $q=5$, by simulating systems containing $N=50q$ particles ($M=50$). The measured threshold values are listed in Table~\ref{TabBinders} and compared to theoretical predictions. 
\begin{table}[ht]
\begin{center}
\begin{tabular}{|c|c|c|c|c|}
\hline
\ $q$ \ & \ $E_{i,j,{\rm min}}^c$ ($\kB T$) \ & \ $\chi^c$ \ & \ $\chi^c$, theory \ & \ ratio \ \\
\hline 
2 & -3.56 & 2.64 & 2 &  1.32 \\
3 &  -3.44 & 3.36 & 2.77 &1.21 \\
4 &  -3.36 & 3.84 & 3.30 &1.16 \\
5 & -3.27 & 4.38 & 3.70 & 1.18 \\
\hline
\end{tabular}
\caption{Numerical threshold values of the energies at contact, $E_{i,j,{\rm min}}$, and the 
Flory parameter $\chi$ as a function of the number of particle families $q$. Theoretical values of $\chi^c$ come from equation~(\ref{chiCtheo}). Here $E_{i,i,{\rm min}}=-4 \kB T$, the system contains $N=50 q$ particles, and the density is $\phi=1/50$.
\label{TabBinders}}
\end{center}
\end{table}
The agreement improves as $q$ grows, thus supporting the theoretical approach of reference~\cite{EPL10} (in principle valid in the large $q$ limit). To ascertain that the measured thresholds were not corrupted by finite size effects, we also simulated a system of $N=210$ particles for $q=3$ ($M=70$). We find $E_{i,j,{\rm min}}^c=-3.47 \kB T$, in satisfactory agreement with the $M=50$ result above. In addition, cluster size distributions for $N=150$ or $N=210$ are undistinguishable. 

\begin{figure}[h]
\begin{center}
\includegraphics[width=0.45\textwidth]{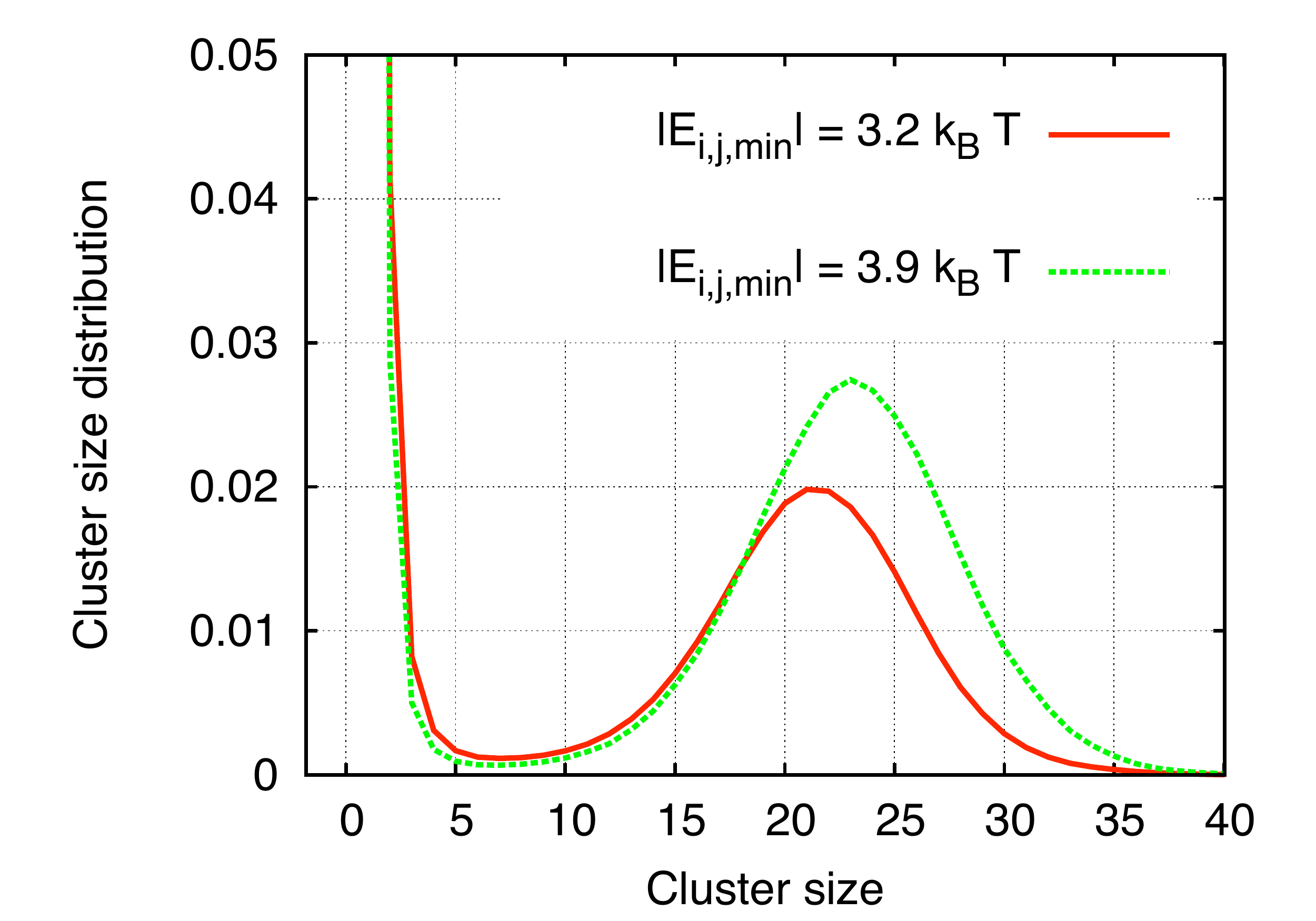} 
\caption{Cluster size distributions $P(k)$. Same parameters as in figure~\ref{snaps2coul}.
\label{dist2coul}}
\end{center}
\end{figure}

Cluster size distributions $P(k)$ are qualitatively unchanged when tuning $|E_{i,j,{\rm min}}|$, as illustrated in figure~\ref{dist2coul} (we get similar results for $q\geq3$). However, when decreasing this parameter or the number of families $q$ at fixed $|E_{i,j,{\rm min}}|$, the typical cluster size decreases because the average short-range attraction decreases. In a similarly way, the fraction of monomers increases slowly when decreasing $|E_{i,j,{\rm min}}|$ (with no apparent singularity at $|E_{i,j,{\rm min}}^c|$). Note that cluster sizes used to compute Binder cumulants above (figure~\ref{binders2coul}) were chosen near the maximum of $P(k)$ to ensure significant statistical sampling.  All these numerical observations confirm our previous theoretical predictions~\cite{EPL10}.

\section{Discussion and conclusion}
\label{discussion}

Our first major goal was to show that cluster phases are robust with respect to the precise shape of interaction potentials, provided that they display both a short-range ($\sim 1$~nm) attraction with a contact energy of a few $\kB T$ and a longer-range repulsion ($\sim 10$~nm), with an energy barrier on the order of $\kB T$ or lower. We have focussed on two limiting potential shapes: a full, unscreened $r^{-4}$ repulsion mediated by the elastic membrane; and a totally screened repulsion, decaying exponentially at long range. In the regime of parameters studied, clusters of size ranging from a few to several tens of particles co-exist with monomers above a critical protein density. We have also demonstrated that taking into account three-body forces propagated by the membrane, which have the same order of magnitude as two-body ones, does not destabilize cluster phases. This suggests that cluster phases are generic in the cell membrane context. 

Our second core objective was to explore how protein diversity affects cluster phases. In this respect,  we have modeled diversity by playing on interaction potentials, both at long and short range. At long range, we have proved that the introduction in the system of a minority ($\lesssim 30$~\%) of up-down symmetric inclusions, experiencing mutual attraction instead of repulsion, does not destabilize cluster phases. At short range, we have confirmed previous analytical calcultations~\cite{EPL10}: modulating interactions favors the segregation of proteins of different families in distinct clusters, provided that proteins of a same family have a slightly higher affinity at contact than proteins of distinct families. Accordingly, the high protein concentration of a membrane ($\sim50$~\% in mass) favors the co-localization of proteins of a same family in same clusters. Were the concentration be too low, proteins would essentially behave as a gas of monomers diffusing freely on the membrane. It would take a long time for them to encounter their partners in the membrane by diffusion. Clusterization and crowding due to high concentration thus favors interactions between them, which should be of biological relevance in terms, e.g., of faithful response to external stimuli~\cite{Gurry09} or of cell adhesion~\cite{Cavey08}.

The present work intends to add some original contributions to a more ambitious program: deciphering cell membrane organization on physical grounds. However, even though we have clarified several points, we have been led to make some assumptions, some of which we intend to discuss now (see also the Supporting Information). The objective of the present work is not to be a definitive answer to the question of membrane organization, but rather to lay the foundations of an emerging scenario.

We have explained why the effective forces propagated by the elastic membrane are dominated at large separations by two- and three-body forces. But sub-dominant, higher-order forces exist that might play a role at intermediate distances and thus influence cluster stability.  However, we conjecture that cluster phases will remain stable in this case, because we have shown that the long-range elastic interaction remain repulsive at large separation. Accordingly, distant clusters will feel a repulsive force that will prevent their further coalescence in larger clusters. To definitely clarify this point, it would be necessary to fully take into account many-body contributions, which requires inverting a $3N \times 3N$ matrix and to calculate its determinant at each Monte Carlo step~\cite{Dommersnes99,Fournier03}. For moderate system sizes, intensive simulations on parallel computers might be able to address this question in a near future.

Even though a live cell is out-of-equilibrium due to energy consumption, experiments done {\em in vitro} after membrane removal from cells still reveal clusters~\cite{Gurry09}. Since in this context, as in the one of artificial vesicles~\cite{Grage11}, no active, energy-consumming processes can limit cluster size, the existence of finite clusters at equilibrium has to be understood. The mechanism explored in this study partakes of this approach. However, cluster phases properties, notably cluster-size distributions, are likely to be modified by active processes. Endocytosis/exocytosis and membrane recycling are likely to enhance clusterization because they fragment large assemblies of proteins. Furthermore, protein conformations change when they are activated~\cite{Salamon00,Alves04}, thus modifying their coupling with the membrane and consequently their interaction parameters. For example, photo-activation of BR in reconstituted proteoliposomes has been shown to promote the formation of clusters~\cite{Kahya02}. Such ingredients could be incorporated in our simulations, e.g. by modeling 
proteins as systems with two internal states, the interaction parameters of which depend on their internal state. The perturbation of equilibrium properties could then be quantified.

To clarify these points, our predictions need to be validated at the experimental level. As already pointed out~\cite{EPL10}, F\"orster Resonance Energy Transfer (FRET)~\cite{FRET} is adapted to quantify the typical distance between fluorescently labeled proteins, and thus their degree of aggregation~\cite{Periole07,Grage11}. By modulating physical parameters such as the inclusion asymmetry, the membrane rigidity or the hydrophobic mismatch~\cite{Botelho06,deMeyer10}, it would in principle be possible to probe this degree of aggregation in model membranes such as vesicles~\cite{Grage11} or stacked supported membranes~\cite{Manghi10}. Fluorescence Correlation Spectroscopy (FCS)~\cite{FRET} should have similar capabilities, since it quantifies the density of entities diffusing in a membrane, a cluster essentially counting as a single entity.

\section*{Supporting Information available}
I. Monte Carlo procedure; II. Poly-dispersity of protein diameters and contact angles; III. On large cluster stability; IV. Comparing two- and three-body interactions for asymmetric inclusions; V. Rotational averaging of elastic interactions for anisotropic inclusions.
 This material is available free of charge via the Internet at \url{http://pubs.acs.org}.

\newpage

\begin{figure}[ht]
\begin{center}
\includegraphics[width=0.65\textwidth]{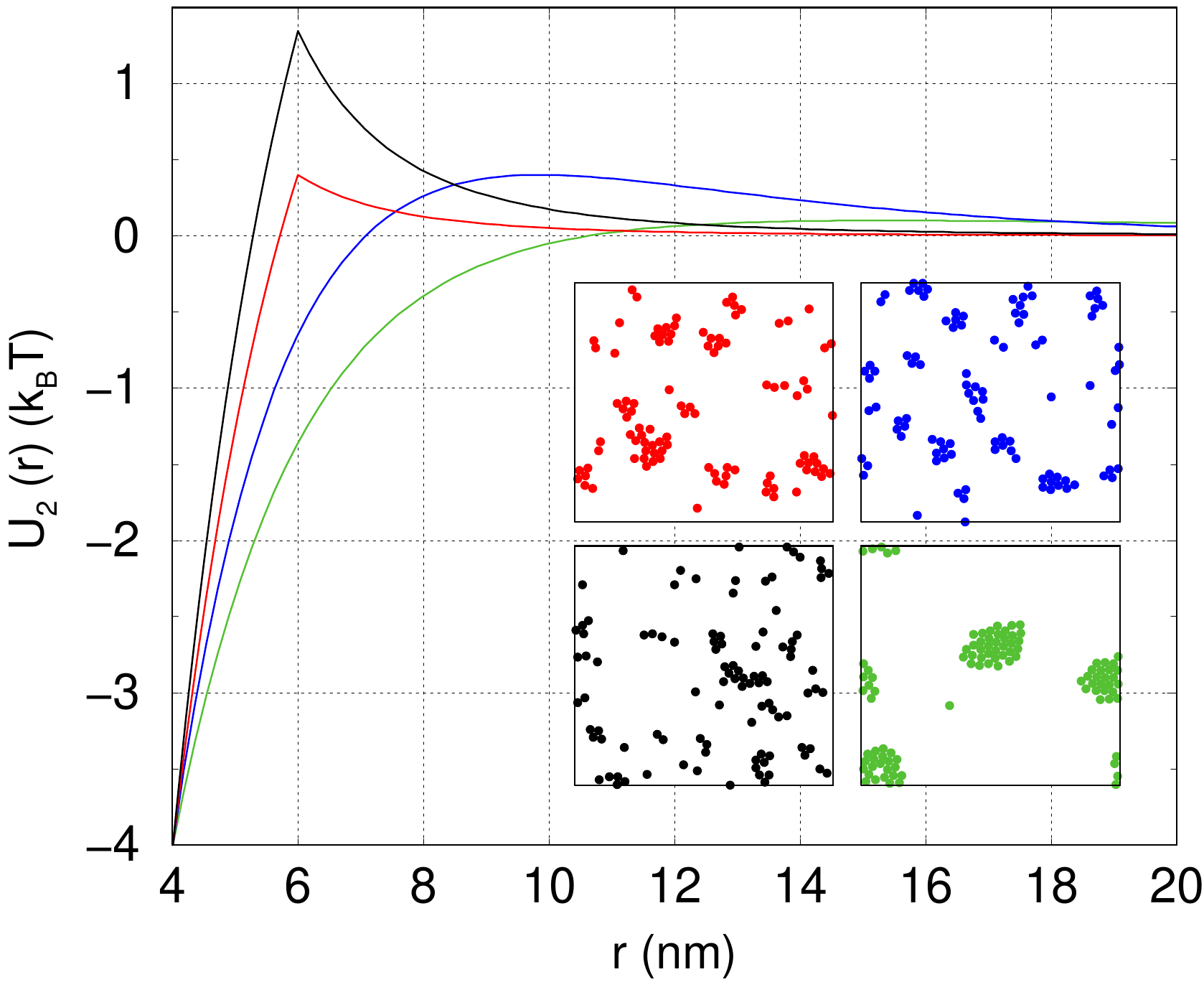}   
\caption{FIG. FOR TOC.}
\end{center}
\end{figure}

\end{document}